\documentclass[amsmath,12pt,floats,floatfix,nofootinbib,tightenlines,showpacs]{revtex4-1}
\usepackage{url}
\usepackage{amsmath}
\usepackage{amssymb}
\usepackage{mathrsfs}
\usepackage{graphicx}
\usepackage{subfigure}
\newcommand{\half}{\frac{1}{2}}
\begin{document}
\title{Sparse spectral-tau method for the
       three-dimensional
       helically reduced wave
       equation on two-center domains}
      \author{{\sc Stephen~R.~Lau}}
      \email[]{lau@math.unm.edu} 
      \affiliation{Department of Mathematics and Statistics\\
          University of New Mexico\\
          Albuquerque, NM 87131.}
      \author{{\sc Richard~H.~Price}}
      \email[]{rprice@phys.utb.edu}
      \affiliation{
          Center for Gravitational Wave Astronomy\\ 
          Department of Physics and Astronomy\\
          University of Texas at Brownsville\\
          Brownsville, TX 78520
              }
\begin{abstract}
We describe a multidomain spectral-tau method for solving the 
three-dimensional helically reduced wave equation on the type 
of two-center domain that arises when modeling compact binary 
objects in astrophysical applications. A global two-center 
domain may arise as the union of Cartesian blocks, cylindrical 
shells, and inner and outer spherical shells. For each such 
subdomain, our key objective is to realize certain (differential 
and multiplication) physical-space operators as matrices acting 
on the corresponding set of modal coefficients. We achieve sparse 
banded realizations through the integration ``preconditioning" 
of Coutsias, Hagstrom, Hesthaven, and Torres. Since ours is the 
first three-dimensional multidomain implementation of the 
technique, we focus on the issue of convergence for the global 
solver, here the alternating Schwarz method accelerated by GMRES. 
Our methods may prove relevant for numerical solution of other 
mixed-type or elliptic problems, and in particular for the 
generation of initial data in general relativity.
\end{abstract}
\maketitle
\newpage
\section{Introduction and preliminaries}\label{sec:introprelim}
\subsection{Introduction}
This paper describes spectral methods designed with a specific 
application in mind: numerical solution of a mixed-type problem 
arising in gravitational physics. In reviewing an ongoing program 
to construct helically symmetric solutions to the Einstein 
equations, this introduction recalls the origins of this 
problem below. However, this paper also serves another 
purpose; it demonstrates that spectral-tau {\em integration 
preconditioning}\footnote{{We use this term to refer to a 
specific technique reviewed below; however, insofar as our work 
is concerned the word {\em preconditioning} might be a misnomer. 
In any case, the technique does achieve {\em sparsification}, and
this is the aspect of the technique we focus on here.}}
yields highly accurate numerical solutions to the helically 
reduced wave equation (HRWE), a mixed-type, variable coefficient, 
linear partial differential equation (PDE) problem, here 
posed on a nontrivial three-dimensional (3D)  domain. 
Ref.~\cite{CHHT} offered spectral-tau integration preconditioning 
as a general-purpose strategy for spectral approximation of 
differential equations, and that reference provides the most 
thorough description and analysis of the technique; related 
techniques were explored in \cite{CHT} (integration 
postconditioning) and \cite{H} (nodal integration preconditioning), 
with applications described in, for example, 
Refs.~\cite{CTdisk,CoutsiasVonWinckel}. However, heretofore, 
spectral-tau integration preconditioning has primarily been 
studied either in the ODE context or for PDE problems posed 
on single and basic two-dimensional (2D) domains (rectangles, 
annuli, and disks), although we have earlier studied a 
2D multidomain scenario \cite{LauPriceI} as 
a warm-up to this work. While the current paper only considers 
the HRWE, a challenging model problem for the aforementioned 
target application, it shows how to implement the technique 
in a 3D multidomain setting, addressing 
several key conditioning issues which would seem to generically 
arise in higher dimensional settings. Therefore, our work 
should facilitate the use of spectral-tau integration 
preconditioning for other elliptic or mixed-type PDE problems. 
We provide more context and a fuller description of these issues
below, but now turn to the physical problem which has motivated 
our work.

The advent of gravitational wave detection has driven 
theoretical studies of gravitational wave sources. A 
source that is possibly interesting for ground-based 
detectors, and perhaps the most exciting source for 
space-based detectors, is the inspiral of two 
comparable mass black holes and their merger to form 
a single black hole.  The early stage of inspiral 
is modeled with reasonable accuracy by perturbations 
of the Newtonian analysis, and the post-merger stage 
can be analyzed with black hole perturbation theory. 
The most difficult stage to analyze is the intermediate 
stage, when a few orbits remain. This epoch of inspiral 
is too late for a modified Newtonian approach, but too 
early for black hole perturbation theory. Yet this is 
the epoch in which a large part of the gravitational 
wave energy is generated.

The importance of, and difficulty in, analyzing this
epoch was the original motivation for an innovative 
approximation, the periodic standing wave (PSW) method. 
This approach uses the fully nonlinear field interactions, 
but models the binary compact objects to be forever on 
circular orbits of constant radius. Therefore, both the 
compact source motions and the fields exhibit helical 
symmetry. Not only does this symmetry reduce the number 
of independent variables, it also completely changes 
the nature of the governing PDEs, turning the 
problem from the hyperbolic evolution of initial data 
to one of mixed-type that is elliptic near a rotation 
axis and hyperbolic well outside the axis and beyond
the orbits in the wave zone of the system.  More details 
of this astrophysical background are given in 
\cite{LauPriceI}. Here we only point out that recent 
supercomputer evolutions of initial black hole binary 
data have been run stably for many orbits 
in the intermediate epoch. See, for example,
Refs.~\cite{Pretorius2005,
UTB2006,
NASA2006,
RIThangup2006,
CalCorn2006,
RITspinflip2007,
Boyle2007,
Jena2008,
CampanelliLoustoZlochower2008,
Bakeretal2008a,
Bakeretal2008b,
Rezzollaetal2008,
CampanelliLoustoNakanoZlochower2009,
SzilagyiLindblomScheel2009,
Gonzalezetal2009,
LoustoZlochower2011,
Lovelaceetal2011} 
(not an exhaustive list),
and \cite{Centrellaetal2010} for a 
recent review. Even the inspiral of binaries with large mass 
ratios \cite{Gonzalezetal2009,LoustoZlochower2011} or high 
spins \cite{Lovelaceetal2011}, both particularly 
challenging cases, can now be computed.
To be sure, recent successes with purely hyperbolic
numerical evolutions have undercut the original motivation
for the PSW approximation. Nevertheless, there remains a 
niche for the PSW approximation for the following reasons. 
First, it should provide a test bench for understanding 
nonlinear gravitational radiation reaction as a local 
process. Second, a helically symmetric solution of
the Einstein equations would be, of its own accord, 
of bewitching interest. 

The numerical computation of PSW fields has, in fact, 
already been carried out, using a single grid and a 
unique method devised especially for the problem by 
one of us (RHP) and coworkers. These computations were 
done in a series of 
steps \cite{Andradeetal2004,BOP2005,BBP2006,BBHP2007,HernandezPrice2009}
moving from linearized scalar fields up to and including the 
nonlinear tensor fields of general relativistic gravity. 
However, the method used proved inherently too limited in accuracy 
to be useful. Furthermore, despite the attractive simplicity
of the computational method, its implementation for general 
relativistic tensor fields proved very challenging.
The astrophysical PSW problem, therefore, can be 
viewed as not yet solved. The spectral methods described 
here are designed to solve this astrophysical 
problem to high accuracy. In any case, as mentioned above, 
our work is relevant as a successful use of spectral-tau 
integration preconditioning for the solution of PDEs 
(even of mixed-type) on nontrivial 3D domains. From this 
standpoint, the astrophysical problem simply provides a 
convenient application, with a particularly interesting 
feature. In the astrophysical problem, the region in which 
the PDEs are hyperbolic ---the distant wave  zone--- is a 
region in which the PDEs have only very small nonlinearities.  
The strong linearities, near the source objects, are confined 
to a region in which the PDEs are elliptic. While we do not
consider nonlinearities in the current paper, the methods 
we introduce for our linear model 
problem deliver sufficient accuracy that nonlinearities can 
almost surely be included.

Multidomain spectral methods for the binary inspiral of 
compact relativistic objects are not new. In pioneering
work, nodal (i.e.~collocation) methods were used by Pfeiffer 
{\em et al.}~\cite{Pfeifferthesis,PKSTelliptic} for the 
elliptical problem of constructing binary black hole initial 
data, and are now being used by the Caltech-Cornell-CITA 
collaboration (see, for example, \cite{SzilagyiLindblomScheel2009}) 
in the fundamentally hyperbolic evolution problem. This work, now 
highly developed, relies on SpEC (the Spectral Einstein Code
\cite{SpEC}), a large C++ project chiefly developed by 
Pfeiffer, Kidder, and Scheel, but also involving many other 
researchers and developers. One certainly might attack the 
problem we consider with that software; in particular with SpEC's 
EllipticModule which uses finite-difference preconditioning
\cite{Orszag,KimParter} and is also already configured to solve 
nonlinear problems. Indeed, the EllipticModule has been used 
to solve the initial value constraint equations on 
essentially the same type of domain we consider 
below\footnote{In fact, the domain decomposition of Pfeiffer 
{\em et al.}~\cite{Pfeifferthesis,PKSTelliptic} motivated
our own choice.}, and we suspect that SpEC could be used to solve 
our problem to high accuracy. In any case, to date
the 3D mixed problem considered here has not been numerically 
solved via spectral methods.

Our previous study \cite{LauPriceI} applied a modal multidomain
spectral-tau method to a model nonlinear problem of two strong
field sources in binary motion with only two spatial dimensions. That
study also relied on integration preconditioning, although the
relevant linear systems were inverted by direct rather than iterative
methods (which was possible since the 2D problem was less memory
intensive). Our 2D study, a proof of concept, showed that high
accuracy could be achieved with relatively modest memory and run-time
requirements.  Here we generalize our 2D method to 3D, that is to
three spatial dimensions and one time dimension, reduced to a problem
with three independent variables by the imposition of helical
symmetry. Due to the larger set of modes needed for the 3D
problem, iterative solution of the relevant linear system is now
necessary. We use the generalized minimum residual method (GMRES)
\cite{GMRES,GolubVanLoan}. Since the amount of work and storage per 
iteration increases with the iteration count \cite{GMRES,GolubVanLoan}, 
preconditioning is a crucial issue (and here we mean further, one might 
even say {\em genuine}, preconditioning on top of the ``integration 
preconditioning"). Through a multilevel preconditioning scheme, we 
will achieve near round-off accuracy for large truncations ($\simeq$ 
600,000 unknowns) with modest iteration counts. Moreover, as we 
achieve a sparse formulation of the relevant linear system, each 
iteration is also fast.
\begin{figure}[t]
\begin{center}
\subfigure[$\;$3d view of domain decomposition.]{
\includegraphics[scale=0.35,clip=true,trim=0 4.25cm 0 2.0cm]{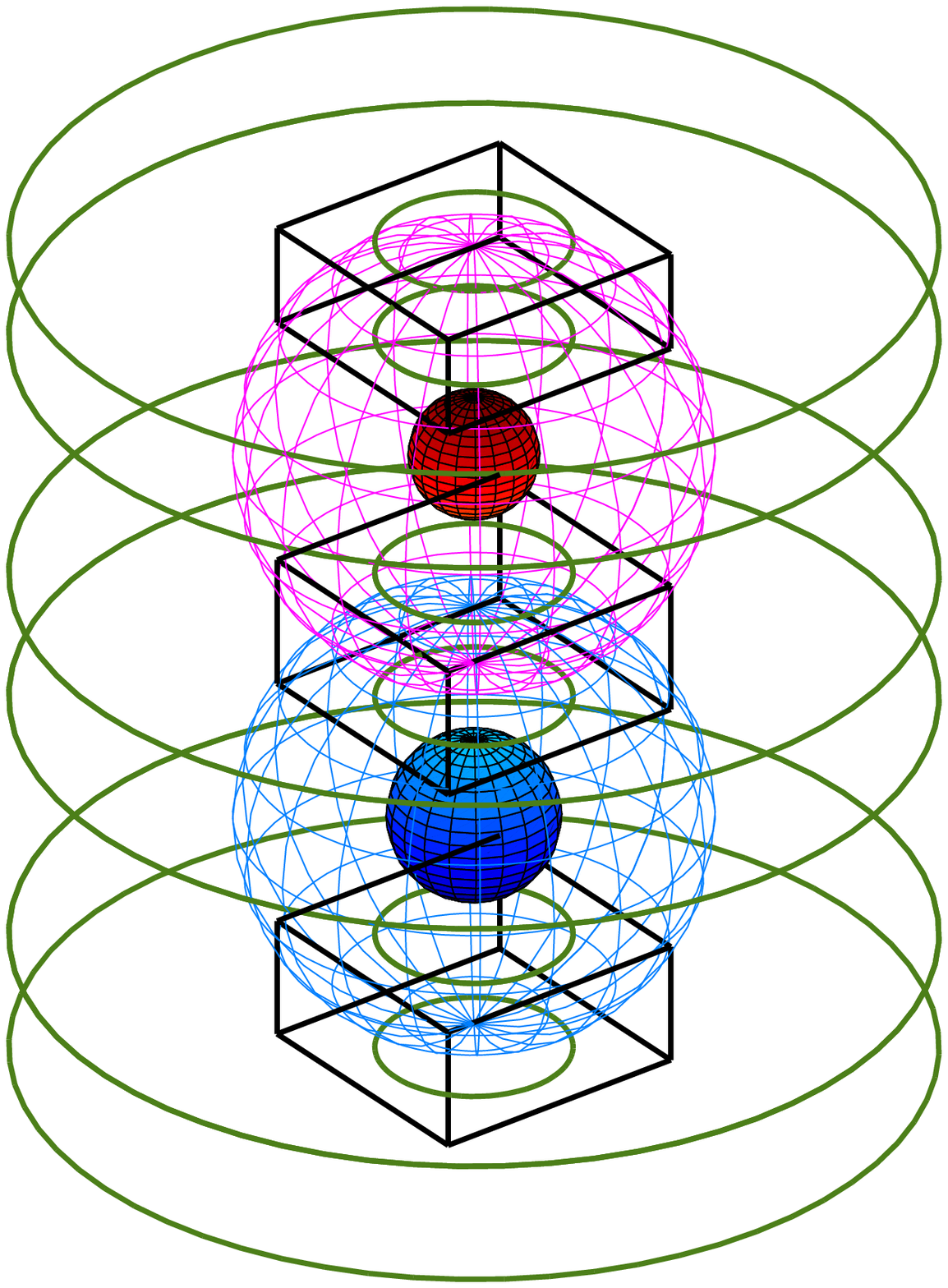}}
\subfigure[$\;$Cross-sectional view.]{
\includegraphics[scale=0.3,clip=true,trim=0 -1.25cm 0 0.0cm]{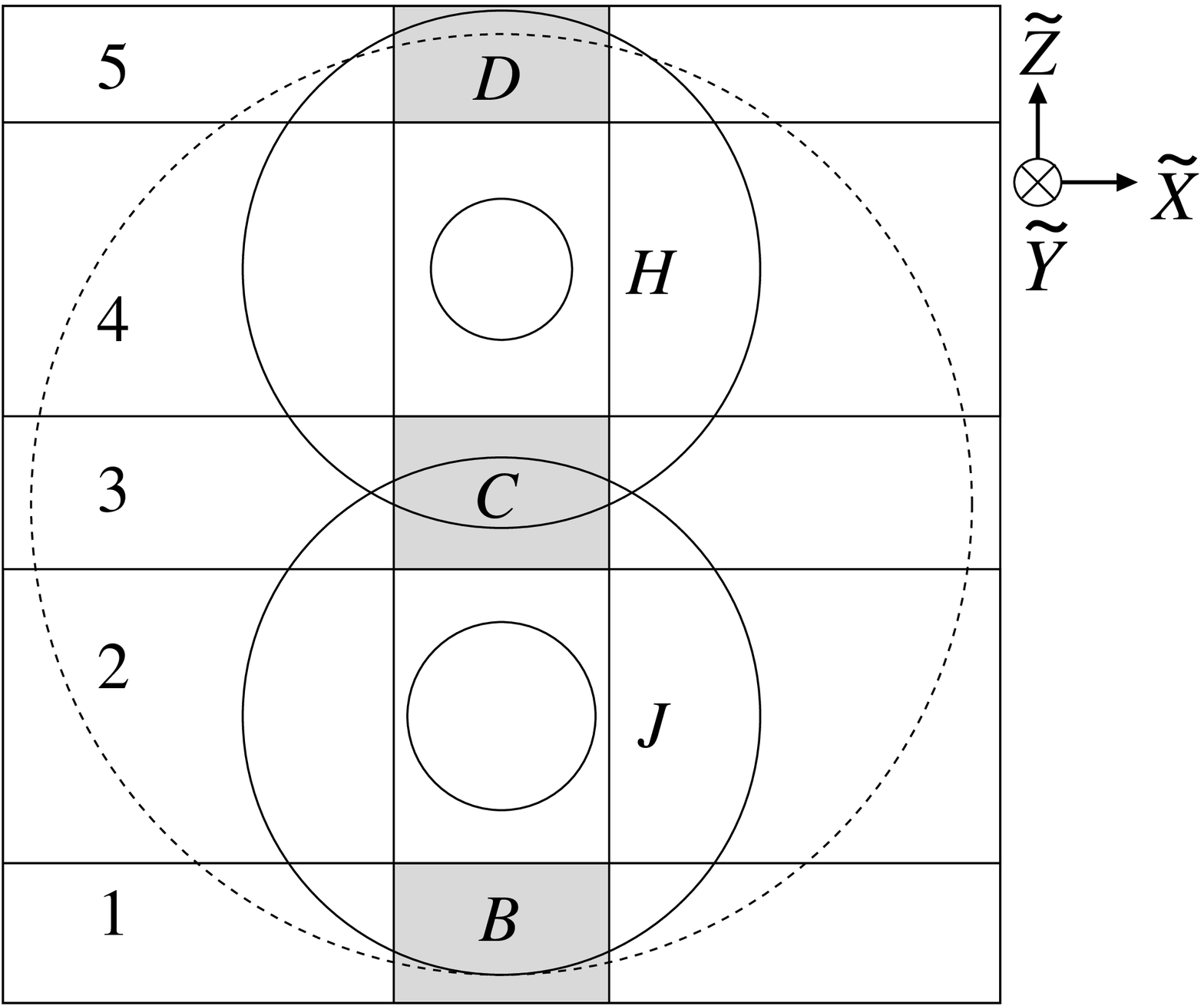}}
\caption{\label{fig:domaindecomp3}
{\sc Domain decomposition.} The
whole inner configuration of 10 subdomains is enclosed within an
outer spherical shell which is not shown, save for its
inner boundary in {\rm (b)}. Our total configuration is
therefore comprised of 11 subdomains.}
\end{center}
\end{figure}

\subsection{Specification of the problem}\label{subsec:domaindecomp}

Before writing down our mixed-type PDE problem, we describe
the two-center (hereafter 2-center) 
domain $\mathcal{D}$ on which the problem is posed, 
first recalling the coordinate conventions of \cite{BOP2005}. Let 
$\{x,y,z\}$ represent the inertial Cartesian system related to 
the spherical-polar system $\{r,\theta,\phi\}$ in the usual 
physicist's convention ($\theta$ and $\phi$ are respectively 
the polar and azimuthal angles). We then introduce a 
``comoving" Cartesian system
\begin{equation}\label{eq:comoving_xyz}
\widetilde{z} = r\cos\theta,\qquad
\widetilde{x} = r\sin\theta\cos(\phi - \Omega t),\qquad
\widetilde{y} = r\sin\theta\sin(\phi - \Omega t),
\end{equation}
where $\Omega < 1$ is a fixed rotation rate. Note that the
$\widetilde{z}$ and $z$-axes coincide, and both are the 
azimuthal axis. Via a simple permutation, we then define
a new comoving system
\begin{equation}\label{eq:tildeXYZ}
\widetilde{X} = \widetilde{y},\qquad 
\widetilde{Y} = \widetilde{z},\qquad
\widetilde{Z} = \widetilde{x},
\end{equation}
for which the $\widetilde{Z}$-axis is not the azimuthal 
axis. If we imagine two compact objects with ``centers" located 
at $\boldsymbol{\xi}_1(t) = a_1\cos(\Omega t)\mathbf{e}_x
+ a_1\sin(\Omega t)\mathbf{e}_y$ and $\boldsymbol{\xi}_2(t) 
= -a_2\cos(\Omega t)\mathbf{e}_x - 
a_2\sin(\Omega t)\mathbf{e}_y$ in the inertial $\{x,y,z\}$ 
system, then the $\widetilde{Z}$-axis connects those compact 
objects. That is, 
$\boldsymbol{\xi}_1 = a_1 \mathbf{e}_{\widetilde{Z}}$ and 
$\boldsymbol{\xi}_2 = -a_2 \mathbf{e}_{\widetilde{Z}}$.
We introduce the coordinates 
$\{\widetilde{r},\widetilde{\theta},\widetilde{\varphi}\}$
as spherical coordinates relative to the comoving
$\{\widetilde{x},\widetilde{y},\widetilde{z}\}$
system. 
We will exclusively work with the {\em comoving} systems 
(or simple translations or polar versions thereof), but
we  will often suppress tildes when doing so will not cause
confusion. We will, for example, use $\{{r},{\theta},{\varphi}\}$, 
hereafter, to mean 
$\{\widetilde{r},\widetilde{\theta},\widetilde{\varphi}\}$; 
these coordinates should not be confused with 
$\{{r},{\theta},{\phi}\}$ of Eq.~(\ref{eq:comoving_xyz}).

Relative to the system $\{\tilde{X},\tilde{Y},\tilde{Z}\}$, the
2-center domain $\mathcal{D}$ that we have used is depicted in
Fig.~\ref{fig:domaindecomp3}.  Topologically, the domain $\mathcal{D}$
is a large solid 3D ball with two excised regions (each a
smaller solid 3D ball). The global domain $\mathcal{D}$ has
been broken into 11 subdomains, each sufficiently simple to allow for
spectral expansions in terms of classical basis functions.  A large
outer shell (labeled $O$ for ``out") is not shown in
Fig.~\ref{fig:domaindecomp3}, but the remaining 10 subdomains which
comprise the ``inner region" are shown. The inner region is comprised
of two ``inner shells" (spherical shells labeled $J$ and $H$), three
``blocks" (rectangular subdomains labeled $B$, $C$, and $D$), and five
``cylinders" (cylindrical shells labeled $1$, $2$, $3$, $4$, and
$5$). Table~\ref{table:domaindecomp3} lists the parameters which 
specify the subdomains comprising $\mathcal{D}$, along with the 
numerical values we have used in the computations to be reported below. 
Its caption describes the relationship between the parameters and the
appropriate comoving system.
\begin{table}\scriptsize
\begin{tabular}{|c|c|c|c|}
\hline
\multicolumn{4}{|c|}{Spherical shells}\\
\hline
\hline
$J$ & $0.4 \leq r \leq 1.1$   & $0 \leq \theta < 2\pi$ & $0 \leq \phi    \leq \pi$\\
$H$ & $0.3 \leq r \leq 1.1$   & $0 \leq \theta < 2\pi$ & $0 \leq \phi    \leq \pi$\\
$O$ & $2.0 \leq r \leq 150.0$ & $0 \leq \theta < 2\pi$ & $0 \leq \varphi \leq \pi$\\
\hline
\hline
\multicolumn{4}{|c|}{Cylindrical shells}\\
\hline
$1$ & $0.452 \leq \rho \leq 2.120$ & $0 \leq \phi < 2\pi$ & $-2.120 \leq \tilde{Z} \leq -1.525$\\
$2$ & $0.452 \leq \rho \leq 2.120$ & $0 \leq \phi < 2\pi$ & $-1.525 \leq \tilde{Z} \leq -0.275$\\
$3$ & $0.452 \leq \rho \leq 2.120$ & $0 \leq \phi < 2\pi$ & $-0.275 \leq \tilde{Z} \leq +0.375$\\
$4$ & $0.452 \leq \rho \leq 2.120$ & $0 \leq \phi < 2\pi$ & $+0.375 \leq \tilde{Z} \leq +1.625$\\
$5$ & $0.452 \leq \rho \leq 2.120$ & $0 \leq \phi < 2\pi$ & $+1.625 \leq \tilde{Z} \leq +2.120$\\
\hline
\hline
\multicolumn{4}{|c|}{Blocks}\\
\hline
$B$ & $-0.640 \leq \tilde{X} \leq 0.640$ & $-0.640 \leq \tilde{Y} \leq 0.640$ &  $-2.120 \leq \tilde{Z} \leq -1.525$\\
$C$ & $-0.640 \leq \tilde{X} \leq 0.640$ & $-0.640 \leq \tilde{Y} \leq 0.640$ &  $-0.275 \leq \tilde{Z} \leq +0.375$\\
$D$ & $-0.640 \leq \tilde{X} \leq 0.640$ & $-0.640 \leq \tilde{Y} \leq 0.640$ &  $+1.625 \leq \tilde{Z} \leq +2.120$\\
\hline
\end{tabular}
\caption{{\em Particular domain decomposition.} The inner shells 
$J$ and $H$ are centered at $(\tilde{X},\tilde{Y},\tilde{Z})
= (0,0,-0.9)$ and $(\tilde{X},\tilde{Y},\tilde{Z}) = (0,0,1.0)$,
and for each shell the polar system $(r,\theta,\phi)$ is relative 
to the Cartesian system arising from {\em translation} of the
$\{\tilde{X},\tilde{Y},\tilde{Z}\}$ system to the shell's
corresponding origin. For each cylinder, the cylindrical system
$(\rho,\phi,\tilde{Z})$ has the standard relationship with the
$\{\tilde{X},\tilde{Y},\tilde{Z}\}$ Cartesian system. Finally,
the outer shell is $O$ centered at the origin of the 
$\{\tilde{X},\tilde{Y},\tilde{Z}\}$ system, but now the 
$(r,\theta,\varphi)$ system is relative to the 
$\{\tilde{x},\tilde{y},\tilde{z}\}$ system.
\label{table:domaindecomp3}
}
\end{table}

The HRWE problem we consider is as follows:
\begin{equation}\label{eq:introHRWE}
L\psi = g    \text{ on } \mathcal{D},\qquad
\psi = h^{-} \text{ on } \partial H^{-} \cup \partial J^{-},\qquad
\left(
\frac{\partial}{\partial r      }
- \Omega 
\frac{\partial}{\partial \varphi}
+ \frac{1}{r}
\right)\psi
= h^{+} \text{ on } \partial O^+,
\end{equation}
where the defining operator is 
\begin{equation}\label{eqn:HRWEoperator}
L 
= \frac{\partial^2}{\partial \tilde{x}^2}
  + \frac{\partial^2}{\partial \tilde{y}^2}
  + \frac{\partial^2}{\partial \tilde{x}^2}
  \,-\, \Omega^2
    \left(\tilde{x}\frac{\partial}{\partial \tilde{y}}
  -       \tilde{y}\frac{\partial}{\partial \tilde{x}}
    \right)^2
= \frac{\partial^2}{\partial\widetilde{X}^2}
+\frac{\partial^2}{\partial\widetilde{Y}^2}
+\frac{\partial^2}{\partial\widetilde{Z}^2}
\,-\,\Omega^2
\left(
\widetilde{Z}\frac{\partial}{\partial\widetilde{X}}
- \widetilde{X}\frac{\partial}{\partial\widetilde{Z}}
\right)^2.
\end{equation}
Here the constant $\Omega$ is the  rotation rate, and $g$ is a prescribed
source. As described in, for example, \cite{LauPriceI} this problem
arises via a helically reduction of the inhomogeneous 3+1 wave
equation (see also the Appendix). Notice that the problem includes
Dirichlet conditions set on the inner boundaries of the spherical
shells $J$ and $H$. The boundary condition set on the outer boundary
of the spherical shell $O$ is of radiative type, and is here expressed
in terms of the polar coordinates $\{r,\theta,\varphi\}$ relative to
$\{\tilde{x},\tilde{y},\tilde{z}\}$. Although this radiation condition
is described precisely below, it may here be thought of as an
inhomogeneous Sommerfeld condition. (The inhomogeneity $h^+$ in 
(\ref{eq:introHRWE}) is a nonlocal expression.) This paper will consider 
only the variable-coefficient {\em linear} problem (\ref{eq:introHRWE}). 
For numerical tests, $g$ is taken as zero on $\mathcal{D}$, but with
distributional support, point sources, located at the centers, 
$\boldsymbol{\xi}_1$ and $\boldsymbol{\xi}_2$, of $J$ and $H$. 
For this choice of $g$, an exact solution is described in the Appendix.  
While we only consider the linear scalar problem (\ref{eq:introHRWE}), 
the helical reduction of the Einstein equations described in 
\cite{BBP2006,BBHP2007} involves a tensorial field resolved into ten 
coupled ``helical scalars" $\psi^{(\alpha\beta)}$ each of which obeys 
a copy $L\psi^{(\alpha\beta)} = g^{(\alpha\beta)}$ of the above
equation. However, for this formulation $g^{(\alpha\beta)}$ is now not
an external source, but rather is a nonlinear coupling function of the
helical scalars built with lower-order terms including some first
derivatives. Therefore, clearly solving the linear problem that we
consider is the first step towards considering the helically reduced
Einstein equations.

\subsection{Overview of 3D spectral-tau integration ``preconditioning"}
\label{subsec:IPCreview}

Mostly focusing on the 3D HWRE in three Cartesian variables
on a rectangular block, this subsection gives a short overview 
of integration preconditioning for spectral-tau methods, in 
particular focusing on the Kronecker product representations 
necessary for 3D. We hope that this overview will provide the 
reader with enough context to follow the heavy details 
encountered later when applying the technique on 3D spherical
and cylindrical shells. Our earlier paper \cite{LauPriceI} 
gave a fuller account of essentially the same issues 
for 2D, many of which change little in going to 3D.
Therefore, in order to here avoid a prohibitively
long discussion, we have opted for a short overview,
and one tied to our particular problem, pointing the reader 
to \cite{LauPriceI} for more details.

The following overview makes use of matrices $D^k$ and 
$B^m_{[n]}$. These respectively represent 
$k$th order 
differentiation and $m$th order integration with respect 
to a basis of Chebyshev polynomials.
As explained later, the subscript $[n]$ 
indicates that the first $n$ rows of a matrix are empty. 
We do not here provide precise expressions for $D^k$ and $B^m_{[n]}$;
however, we list the following key properties exploited later: (i)
$D^k$ is dense upper triangular, (ii) $B^n_{[n]}$ is sparse and 
banded with upper and lower bandwidth $n$, and (iii) 
$B^n_{[n]} D^k = B^{n-k}_{[n]}$ for
$n\geq k$. Here $B^0_{[n]} \equiv I_{[n]}$ is the identity matrix,
except that each entry in its first $n$ rows is a $0$. Our earlier
paper \cite{LauPriceI} gave the precise expressions for $D\equiv D^1$,
$D^2$, $B_{[1]}\equiv B^1_{[1]}$, and $B^2_{[2]}$; our treatment
of the 3D HRWE (a second order equation) only requires these matrices.
That reference also discusses the necessary rescalings of these
matrices for work with an arbitrary interval rather than the
standard interval $[-1,1]$ for Chebyshev polynomials. Of course,
Ref.~\cite{CHHT} also considered such expressions and identities, even
for more general basis functions.

\subsubsection{Direct product representations}
A function on any of our 3D subdomains is encoded by the modal 
coefficients for its spectral expansion, and this set of modal 
coefficients is often here viewed as a direct (Kronecker) 
product. For example, let us consider a rectangular block 
delineated by the above comoving coordinates 
$\{\widetilde{X},\widetilde{Y},\widetilde{Z}\}$, but for 
the rest of this overview let us suppress the tildes on these 
coordinates. Suppose a function $\psi(X,Y,Z)$ on the 
block is formally represented as a triple Chebyshev expansion
\begin{equation}
\psi(X,Y,Z) = \sum_{n=0}^\infty
              \sum_{m=0}^\infty
              \sum_{p=0}^\infty
              \widetilde{\psi}_{nmp}
              T_n(\xi(X))T_m(\eta(Y))T_p(\chi(Z)),
\end{equation}
where $(\xi(X),\eta(Y),\chi(Z))$ maps our block to the standard
cube $[-1,1]^3$. To get an approximation of $\psi(X,Y,Z)$, 
we consider the truncated series
\begin{equation}
\mathcal{P}_{N_X,N_Y,N_Z}\psi(X,Y,Z) =
              \sum_{n=0}^{N_X}
              \sum_{m=0}^{N_Y}
              \sum_{p=0}^{N_Z}
              \widetilde{\psi}_{nmp}
              T_n(\xi(X))T_m(\eta(Y))T_p(\chi(Z)),
\end{equation}
so that $\psi(X,Y,Z)$ is represented (either exactly or 
approximately) by a three-index set 
$\{\widetilde{\psi}_{nmp}: 
0 \leq n \leq N_X,
0 \leq m \leq N_Y,
0 \leq p \leq N_Z\}$ of modal coefficients.
We represent this finite collection of modal coefficients 
as a column vector $\widetilde{\boldsymbol{\psi}}$, with 
components $\widetilde{\boldsymbol{\psi}}(\alpha) =
(\widetilde{\boldsymbol{\psi}})_\alpha$ determined by the
direct product representation\footnote{We could have instead
taken
$$  
\widetilde{\boldsymbol{\psi}}(m(N_X+1)(N_Z+1)+p(N_X+1)+n)=
\widetilde{\psi}_{nmp},
$$
which might prove advantageous for representation of the
$\Omega^2$ term in (\ref{eqn:block_ibp}). Our choice
(\ref{eq:block_dprep}) has been determined by technical
decisions made during the initial construction of our 
code. In any case, based on some experimentation, we 
believe this choice makes little difference, at least for 
$\Omega \lesssim 0.5$ (well in the range of rotation 
rates we aspire to treat).}
\begin{equation}\label{eq:block_dprep}
\widetilde{\boldsymbol{\psi}}(n(N_Y+1)(N_Z+1)+m(N_Z+1)+p)=
\widetilde{\psi}_{nmp}.
\end{equation}
A single matrix operating on the vector $\widetilde{\boldsymbol{\psi}}$
(all modal coefficients representing the given function) may 
then equivalently be considered as having block-elements which 
are other matrices. We always view the modal set for a function 
on a cylindrical or rectangular subdomain as a direct product 
of {\em three} one-dimensional sets. However, in the case of the 
spherical shells ($J$, $H$, and $O$), we sometimes view the set 
of modal coefficients as the direct product of only {\em two} 
sets, the set corresponding to the radial modes and the set 
corresponding to the spherical harmonic modes (which involve both 
the polar and azimuthal angles). 

In our notation, operators corresponding to a single dimension, that 
is ``simple'' matrices (whose elements are numbers, not matrices), 
are usually represented by a capital in an ordinary font, such as 
the identity operator/matrix $I_X$ or the matrix $D_Z$ which realizes 
differentiation by $Z$. Matrices which act on the full set of 
modal coefficients
are represented by a calligraphic capital, for example $\mathcal{B}$. 
Thus, if $B_{Z[1]}$ represents integration in $Z$, then on a rectangular 
subdomain we might have ${\mathcal B}=I_X\otimes I_Y\otimes B_{Z[1]}$ 
as the matrix which, when applied to a vector 
$\widetilde{\boldsymbol{\psi}}$ holding the full set of modal 
coefficients, realizes integration in $Z$ with no action in $X$ 
or $Y$. That is, if $\psi(X,Y,Z)$ has modal coefficients 
$\widetilde{\boldsymbol{\psi}}$, then formally $\int \psi(X,Y,Z)dZ$ 
has modal coefficients $\mathcal{B}\widetilde{\boldsymbol{\psi}}$. 
The $[1]$ on $B_{Z[1]}$ indicates that all entries in the first row (in
fact, the zeroth row by our conventions) of this matrix are zero, so 
that $(\mathcal{B}\widetilde{\boldsymbol{\psi}})(\alpha) = 0$ whenever 
$p = 0$ for the index $\alpha = n(N_Y+1)(N_Z+1)+m(N_Z+1)+p$\;
[cf.~Eq.~\eqref {eq:block_dprep}]. This choice would 
fix the integration constant (a function of $X$ and $Y$) in 
$\int \psi(X,Y,Z)dZ$, but this empty row might also be subsequently 
filled with a ``tau-condition," that is another vector chosen to 
fix a different constant.

\subsubsection{Integration preconditioning}
Let us briefly review the key ideas behind the technique of
integration preconditioning, continuing to assume a rectangular block
subdomain and also assuming the operator (\ref{eqn:HRWEoperator}) for
the HRWE. (See Refs.~\cite{CHHT,LauPriceI} for nonlinear scenarios, \cite{CHHT}
for more complicated 1D operators, and Refs.~\cite{CHHT,HGG} for
more exotic basis functions). After enough invocations of the Leibniz
rule, we may express the operator in (\ref{eqn:HRWEoperator})
(again with tildes 
suppressed) as
\begin{equation}\label{eqn:block_ibp}
L = \partial^2_Y
+   \partial^2_X(1 - \Omega^2 Z^2)
+   \partial^2_Z(1 - \Omega^2 X^2)
   -\Omega^2 (\partial_X X
             +\partial_Z Z
             -2\partial_X\partial_Z XZ).
\end{equation}
We view this equation as an {\em operator identity}, that 
is the partial derivatives see both terms like $Z^2$ and
$XZ$ to the right, and also the function (not shown) on 
which $L$ will eventually act. The Chebyshev polynomials 
$T_n(\xi)$ obey the three-term recurrence 
$2\xi T_n(\xi) = T_{n+1}(\xi) + T_{n-1}(\xi)$. 
Here $\xi$ may be viewed as a suitable rescaling of 
either $X$, $Y$, or $Z$. Therefore, multiplication by 
the independent variable (here $\xi$) is represented in 
the corresponding space of modal coefficients by a banded 
(evidently a tridiagonal) matrix $A_\xi$. 
In fact, multiplication by a polynomial $p(\xi)$ is similarly 
represented by a banded matrix $p(A_\xi)$. Now, the matrix 
which represents $L$ is
\begin{align}
\mathcal{L} & =
      I_X \otimes D^2_Y \otimes I_Z
  +   D^2_X \otimes I_Y \otimes (I_Z - \Omega^2 A_Z^2)
  +   (I_X - \Omega^2 A_X^2) \otimes I_Y \otimes D^2_Z
\nonumber \\
& - \Omega^2 (
      D_X A_X \otimes I_Y \otimes I_Z
    + I_X \otimes I_Y \otimes D_Z A_Z
    - 2D_X A_X \otimes I_Y \otimes D_Z A_Z
            ).
\end{align}
where each $D$ represents differentiation in the space of modal 
coefficients for one coordinate. This matrix is of the general
form
\begin{equation}\label{eq:generalL}
\mathcal{L}
= \sum_{i=0}^{2}\sum_{j=0}^{2}\sum_{k=0}^{2}
D_X^i \otimes D_Y^j \otimes D_Z^k
\left(
\sum_{r=0}^{2}\sum_{s=0}^{2}\sum_{t=0}^{2}
p_{ijk,rst} A_X^r \otimes A_Y^s \otimes A_Z^t
\right),
\end{equation}
where the $p_{ijk,rst}$ are constants (most zero in our case).
In Eq.~(\ref{eq:generalL}) the matrix within the parenthesis 
is banded and sparse; however, overall $\mathcal{L}$ is neither, 
since these desirable features are spoiled by the derivative 
matrices (see the second paragraph of this subsection).

The idea behind integration preconditioning is 
to ``undo" all of the matrix differentiations which appear 
in (\ref{eq:generalL}) through repeated application of 
integration matrices [cf.~point (iii) in the second
paragraph of this subsection]. To illustrate, we 
consider the modal representation
$\mathcal{L}\widetilde{\boldsymbol{\psi}} = 
\widetilde{\boldsymbol{g}}$
of (\ref{eq:introHRWE}) on the rectangular block, ignoring 
for the time being the issue of boundary conditions. 
Introducing $\mathcal{B} \equiv B^{2}_{X[2]} \otimes
B^{2}_{Y[2]} \otimes B^{2}_{Z[2]}$, we then form
$\mathcal{B}\mathcal{L}\widetilde{\boldsymbol{\psi}} 
= \mathcal{B}\widetilde{\boldsymbol{g}}$. The 
coefficient matrix of the new ``preconditioned" 
system is then
\begin{align}\label{eq:blockBL}
&\mathcal{BL} =
B_{X[2]}^2 \otimes I_{Y[2]} \otimes B_{Z[2]}^2
\nonumber \\
& + I_{X[2]} \otimes B_{Y[2]}^2 \otimes 
    (B_{Z[2]}^2 - \Omega^2 B_{Z[2]}^2 A_Z^2)
  + (B_{X[2]}^2 - \Omega^2 B_{X[2]}^2 A_X^2) \otimes 
     B_{Y[2]}^2 \otimes I_{Z[2]}
\nonumber \\
& -\Omega^2 (
      B_{X[2]} A_X \otimes B_{Y[2]}^2 \otimes B_{Z[2]}^2
    + B_{X[2]}^2 \otimes B_{Y[2]}^2 \otimes B_{Z[2]} A_Z
    - 2B_{X[2]} A_X \otimes B_{Y[2]}^2 \otimes B_{Z[2]} A_Z
            ).
\end{align}
Because it is built only with $B$s and $A$s, this matrix is 
sparse and banded, albeit with large bandwidth due to the direct 
product structure. 

The matrix $\mathcal{B}\mathcal{L}$ has many empty rows, signaling
missing information. The spectral-tau procedure is to put the
``tau conditions,'' here the boundary conditions, in these empty 
rows, and the corresponding inhomogeneous values in 
$\mathcal{B}\widetilde{\boldsymbol{g}}$.
When this procedure is carried out correctly, with 
due regard to possible repetition in the specification 
of boundary data, the empty rows provide precisely the 
space needed for the boundary data of a well-posed problem.
To enforce boundary conditions for the example at hand, we 
proceed as follows. Define, for example, $h^{+}(X,Y) = 
\psi(X,Y,Z_\mathrm{max})$ and $h^{-}(X,Y) =
\psi(X,Y,Z_\mathrm{min})$. Then Dirichlet
boundary conditions along the $XY$-faces of a block
are expressible as
\begin{equation}\label{eq:XYfaceBC}
\sum_{p=0}^{N_Z} \widetilde{\psi}_{nmp} \delta_p^{\pm}
= \tilde{h}^\pm_{nm},
\end{equation}
where a double Chebyshev projection appears on the right-hand
side. Moreover, $\delta^+$ (all 1's) and $\delta^-$ 
(alternating $+1$ and $-1$) are the $(N_Z+1)$ dimensional
``Dirichlet vectors.''
Similar equations correspond to $YZ$ and $XY$ faces,
and in all $2(N_X+1)(N_Y+1)+2(N_Y+1)(N_Z+1)+2(N_X+1)(N_Z+1)$
such equations are possible. However, there are only
$$
2(N_X+1)(N_Y+1)+2(N_Y+1)(N_Z+1)+2(N_X+1)(N_Z+1)
- 4(N_X+N_Y+N_Z+1)
$$
available empty rows in (\ref{eq:blockBL}). However, there
are precisely $4(N_X+N_Y+N_Z+1)$ linear dependencies amongst
the set of all possible boundary equations, owing to the fact
that faces share common edge values. Table \ref{tab:blockrows}
gives our prescription for filling empty rows.
\begin{table}
\begin{tabular}{|l|l|l|}
\hline
\multicolumn{1}{|c|}{Face} &
\multicolumn{1}{|c|}{Rows} &
\multicolumn{1}{|c|}{Index restrictions} \\
\hline
$Z = Z_\mathrm{min}$ & $n(N_Y+1)(N_Z+1)+m(N_Z+1)+0$ & $0\leq n\leq N_X,\; 0\leq m\leq N_Y$ \\
$Z = Z_\mathrm{max}$ & $n(N_Y+1)(N_Z+1)+m(N_Z+1)+1$ & $0\leq n\leq N_X,\; 0\leq m\leq N_Y$ \\
\hline
$Y = Y_\mathrm{min}$ & $n(N_Y+1)(N_Z+1)+p$         & $0\leq n\leq N_X,\; 2\leq p\leq N_Z$\\
$Y = Y_\mathrm{max}$ & $n(N_Y+1)(N_Z+1)+(N_Z+1)+p$ & $0\leq n\leq N_X,\; 2\leq p\leq N_Z$ \\
\hline
$X = X_\mathrm{min}$ & $m(N_Z+1)+p$                & $2\leq m\leq N_Y,\; 2\leq p\leq N_Z$\\
$X = X_\mathrm{max}$ & $(N_Y+1)(N_Z+1)+m(N_Z+1)+p$ & $2\leq m\leq N_Y,\; 2\leq p\leq N_Z$\\
\hline
\end{tabular}
\caption{{\sc Filling of empty rows for blocks.}
\label{tab:blockrows}
}
\end{table}

As a result of the integration preconditioning, we have reformulated 
the set of equations in terms of matrices with a drastic reduction 
in nonzero elements. In the context of ODEs, that is in the 1D 
origins of this method, Ref.~\cite{CHHT} has thoroughly studied the 
condition number of the resulting preconditioned matrix with respect 
to norms that arise from diagonal equilibration. While in 
ODE settings integration preconditioning often improves the 
conditioning of the original system, in the PDE context at hand 
$B^{2}_{X[2]} \otimes B^{2}_{Y[2]} \otimes B^{2}_{Z[2]}$ is not an 
optimal preconditioner in that it does not approximate (in any 
measure that we are aware of) the inverse of the original coefficient 
matrix. Nevertheless, one should expect that the ``preconditioned" 
coefficient matrix has a more clustered spectrum, since the $B$s are 
compact operators (even as infinite dimensional matrices). A 
clustered spectrum often yields favorable convergence properties 
in the context of Krylov iterative methods. Regardless, 
{\em sparsification} is a desirable property, since it clearly 
affords a {\em fast} matrix-vector multiply in Krylov methods. 
Therefore, for multidimensional problems we are more comfortable 
focusing on the sparsifying aspect of the technique, with the 
understanding that {\em further preconditioning} (described below) 
on top of the ``integration preconditioning" will be required to 
enhance convergence of the underlying linear solver (in our case
GMRES).

\section{Sparse spectral approximation of the 3d HRWE}
\label{sec:sparse3dHRWE}
Starting with one of the forms appearing in 
(\ref{eqn:HRWEoperator}) and after further transformations, 
this section describes the spectral-tau representation of $L$ 
on each of the basic subdomains, except for the block case 
which we have already described in Sec.~\ref{subsec:IPCreview}. 
These descriptions allow for implementation of the lefthand 
side of (\ref{eq:introHRWE}) as a ``matrix-vector multiply," 
an implementation required by the iterative solver GMRES 
\cite{GMRES}. 

\subsection{Outer spherical shell}\label{subsec:outershell}
In the polar system associated with the comoving system 
(\ref{eq:comoving_xyz}) the operator (\ref{eqn:HRWEoperator}) 
becomes
\begin{equation}\label{eq:r2LonO}
r^2 L = r^2\Delta - \Omega^2 J_O,
\qquad
r^2\Delta = \partial_r^2 r^2 - 2 \partial_r r + \Delta_{S^2},
\qquad 
J_O = r^2\partial_\varphi^2,
\end{equation}
where $\Delta_{S^2}$ is the unit-sphere Laplacian and $O$ stands for
the {\em outer} spherical shell. These equations
should be viewed as operator identities acting on scalar functions. 
The solution $\psi$ to $L\psi = g$ is formally represented as the 
triple series
\begin{align}
\psi(r,\theta,\varphi) & = \sum_{n=0}^\infty 
                        \sum_{\ell = 0}^\infty 
                        \widetilde{\psi}_{\ell 0 n} 
                        \overline{P}_{\ell 0}(\cos\theta) 
                        T_n(\xi(r))
                        \nonumber \\
                    & + \sum_{n=0}^\infty 
                        \sum_{\ell = 1}^\infty  
                        \sum_{m=1}^\ell 
                        \overline{P}_{\ell m}(\cos\theta)
                        \big[
                             \widetilde{\psi}_{\ell, 2m-1, n}
                             \cos (m\varphi)
                           + \widetilde{\psi}_{\ell, 2m, n}
                             \sin (m\varphi)
                        \big]
                        T_n(\xi(r)),
\end{align}
where the $\overline{P}_{\ell m}(\cos\theta)$ are normalized 
associated Legendre functions \cite{AS} and 
$\xi(r)$ maps the radial domain to the standard interval $[-1,1]$.
The corresponding numerical approximation is the following 
finite expansion:
\begin{align}
\mathcal{P}_{N_r,N_\theta}
\psi(r,\theta,\varphi) & = \sum_{n=0}^{N_r} 
                        \sum_{\ell = 0}^{N_\theta}
                        \widetilde{\psi}_{\ell 0 n}
                        \overline{P}_{\ell 0}(\cos\theta)
                        T_n(\xi(r))
                        \nonumber \\
                    & + \sum_{n=0}^{N_r}
                        \sum_{\ell = 1}^{N_\theta}  
                        \sum_{m=1}^{N_\theta}
                        \overline{P}_{\ell m}(\cos\theta)
                        \big[
                             \widetilde{\psi}_{\ell, 2m-1, n}
                             \cos (m\varphi)
                           + \widetilde{\psi}_{\ell, 2m, n}
                             \sin (m\varphi)
                        \big]
                        T_n(\xi(r)).
\label{eqn:sphchbtrunc}
\end{align}

We represent the triply-indexed modal coefficients
$\widetilde{\psi}_{\ell q n}$ as a vector $\widetilde{\psi}(\alpha)$
of length $(N_\theta+1)(2N_\theta+1)(N_r+1)$, with the 
two notations connected by
\begin{equation}\label{eq:shellvector}
\widetilde{\boldsymbol{\psi}}(\ell(2N_\theta+1)(N_r+1)+q(N_r+1)+n)
= \widetilde{\psi}_{\ell q n},
\end{equation}
For $\ell < N_\theta$ the second sum over $m$ in
(\ref{eqn:sphchbtrunc}) includes too many
terms. Indeed, $m$ should run from $1$ to $\ell$ only 
(with the $m=0$ terms appearing in the first sum); therefore,
whenever $q > 2\ell$, we must set $\widetilde{\psi}_{\ell q n} = 0$ by
hand. We have enlarged the space of modal
coefficients for later convenience when using spherical harmonic
transformations. With this remark in mind, for our 
representation \eqref{eq:shellvector} the index $\alpha$ of the
vector $\widetilde{\psi}(\alpha)$ starts at $0$ and takes on 
all values corresponding to the ranges $0 \leq \ell \leq N_\theta$, 
$0 \leq q \leq 2N_\theta$, and $0 \leq n \leq N_r$. We denote
by $\mathbb{P}$ the projection matrix whose range is the set 
of vectors associated with proper spherical harmonic expansions,
\begin{equation}\label{eq:shellvectorsubspace}
(\mathbb{P}\widetilde{\boldsymbol{\psi}})
(\ell(2N_\theta+1)(N_r+1)+q(N_r+1)+n)
= 0, \text{ for } q > 2\ell.
\end{equation}

Let us first consider a sparse approximation of the Laplacian
term $r^2\Delta$, which
from (\ref{eq:r2LonO}) has the spectral representation
\begin{equation}
\mathcal{A}_r^2\boldsymbol{\Delta} = 
       \mathbb{P}\big[ I_\theta \otimes I_\varphi
       \otimes ({D}_r^2 A_r^2 
     - 2 {D}_r A_r)
     - \mathscr{L}^2 \otimes I_r],
\end{equation}
where $\mathcal{A}_r^2 = I_\theta \otimes I_\varphi \otimes A_r^2$,
and $A_r$ is the matrix equivalent to multiplying $r$-dependent
functions by a factor of $r$. 
In the first term 
within the square brackets $I_\theta \otimes I_\varphi\otimes$ 
means that there are no operations mixing modes 
$\widetilde{\psi}_{\ell q n}$ with different values of
$\ell$, or of $q$
(i.e.~the dual indices to $\theta$ and $\phi$).  The operator 
$(D_r^2 A_r^2 - 2 D_r A_r)$ is the matrix 
equivalent of the partial differentiation 
$\partial_r^2 r^2 - 2 \partial_r r $ in
(\ref{eq:r2LonO}).
The matrix $\mathscr{L}^2$, representing $-\Delta_{S^2}$
in (\ref{eq:r2LonO}), is comprised of $(N_\theta+1)$
constant blocks $\ell(\ell+1)
I_{(2N_\theta+1)\times 
   (2N_\theta+1)}$ in 
each subspace labeled by $\ell$.

To get a sparse form of the Laplacian, we define 
$\mathcal{B} = I_\theta \otimes I_\varphi \otimes B_r^2$
and write the expression
\begin{equation}\label{eq:precondLapO}
(\mathcal{B}\mathcal{A}_r^2\boldsymbol{\Delta})^\mathrm{modified} 
= \mathbb{P}\big[I_\theta \otimes I_\varphi 
\otimes (I_{r[2]} A_r^2 - 2 B_{r[2]} A_r)  
+ \mathscr{L}^2 \otimes B_{r[2]}^2\big]
+ (I_\theta \otimes I_\varphi 
\otimes I_r - \mathbb{P})\,.
\end{equation}
Here the ``modified'' notation indicates that, by the addition of the
last term above, 1's have been placed on the diagonal in 
rows set to zero by the projection operation, so that the result 
is a nonsingular matrix. Therefore, to ensure that a solution
$\widetilde{\boldsymbol{\psi}}$ to the corresponding linear 
system obeys
\begin{equation}
\widetilde{\boldsymbol{\psi}}(\ell(2N_\theta+1)(N_r+1)+q(N_r+1)+n)
= 0,\qquad \mathrm{for}\; q > 2\ell\,.
\end{equation}
We demand that the source obeys $\widetilde{\boldsymbol{g}} =
\mathbb{P}\widetilde{\boldsymbol{g}}$.  Finally, from
(\ref{eq:r2LonO}) the sparse preconditioned form of 
the operator $J_O$ is
\begin{equation}
\mathcal{B} \mathcal{J}_O = 
-\mathbb{P}\big[I_\theta \otimes 
\mathscr{M}^2 \otimes B_{r[2]}^2 A_r^2\big],
\end{equation}
where $\mathscr{M}^2
= \mathrm{diag}(0,1,1,4,4,\cdots,N_\theta^2,N_\theta^2)$ is 
the $(2N_\theta+1)$-by-$(2N_\theta+1)$ matrix representing
$-\partial_\varphi^2$. Therefore,
$(\mathcal{B}\mathcal{A}_r^2\boldsymbol{\Delta})^\mathrm{modified} -
\Omega^2 \mathcal{B}\mathcal{J}_O$ is our sparse form of the overall
coefficient matrix, prior to inclusion of boundary conditions.

We now consider specification of outer radiation conditions, for which
we summarize results given in \cite{LauPriceI}. Specification of
Dirichlet conditions on the inner boundary $\partial O^-$ of the
outer shell $O$ is essentially the same as specification on the
boundaries $\partial H^\pm$ of the inner shell $H$, and we describe
that specification in detail below.  The specification at $\partial
O^+$, however, involves radiation conditions. To define these, we set
$R = r_\mathrm{max}$, the radial coordinate value of $\partial O^+$,
and introduce 
\begin{equation}\label{Vnudef}
V_{\ell + 1/2}(\xi) =
          \sqrt{\frac{\pi \xi}{2}}
          \exp\big[-\mathrm{i}
          \big(\xi - {\textstyle\frac{1}{2}} \pi \kappa
        - {\textstyle \frac{1}{4}}\pi\big)\big]
          H^{(+)}_{\ell + 1/2} (\xi),
\end{equation}
which satisfies $V_{\ell + 1/2}(\xi)\sim 1$ as $\xi \rightarrow\infty$. 
Here $H^{(+)}_{\ell + 1/2}(\xi)$ is the cylindrical Hankel function of 
the first kind, of  
half-integer order $\ell + 1/2$. For our radiative boundary 
condition we will need the 
``frequency--domain kernel,''
\begin{equation}
v_{\ell + 1/2}(\xi) \equiv
\xi\, \frac{V'_{\ell + 1/2}(\xi)}{V_{\ell + 1/2}(\xi)}\,,
\end{equation}
which is computable as a continued fraction via Steed's
algorithm \cite{ThomBarn}. Radiation conditions involve
\begin{equation}
p \equiv m\Omega R +\mathrm{Im}\left(  v_{\ell+1/2}(m\Omega R)\right),
\quad
q \equiv 1 - \mathrm{Re}\left( v_{\ell+1/2}(m\Omega R)\right),
\end{equation}
with $p = 0$ and $q = \ell + 1$ for $m = 0$ modes 
(see Ref.~\cite{LauPriceI} for details). 
The $p$ and $q$ here (in particular the $q$) are not related 
to the indices on 
$\widetilde{\psi}_{\ell q n}$. Both uses of $p$ and $q$
will not appear in the same formula. As tau-conditions, 
our radiation conditions are then expressible as
\begin{subequations}\label{eq:rmaxuBCs}
\begin{align}
\sum_{n=0}^{N_r}\big(
R \widetilde{\psi}_{\ell, 2m, n} \nu^+_n
+ p \widetilde{\psi}_{\ell, 2m-1, n} \delta^+_n
+ q \widetilde{\psi}_{\ell, 2m, n} \delta^+_n\big)  & = 0\\
\sum_{n=0}^{N_r}\big(
R \widetilde{\psi}_{\ell, 2m-1, n} \nu^+_n
- p \widetilde{\psi}_{\ell, 2m, n} \delta^+_n
+ q \widetilde{\psi}_{\ell, 2m-1, n} \delta^+_n\big) & = 0\,.
\end{align}   
\end{subequations}
Here $\delta^+$ (all 1's) and $\delta^-$ (alternating $+1$ 
and $-1$) are the $(N_r + 1)$ dimensional ``Dirichlet vectors'' 
used to impose Dirichlet conditions at the endpoints of a 
coordinate range. Similarly, $\nu^+$ and $\nu^-$ are the 
$(N_r+1)$ dimensional ``Neumann vectors'' used to impose 
derivative conditions at the endpoints. Details are given 
in \cite{CHHT,LauPriceI}.

Along the block-diagonal of the coefficient matrix 
$(\mathcal{B}\mathcal{A}_r^2\boldsymbol{\Delta})^\mathrm{modified} 
- \Omega^2 \mathcal{B}\mathcal{J}_O$, there are 
$(N_r+1)$-by-$(N_r + 1)$ blocks, one for 
each $(\ell,q)$ pair. When $q$ exceeds $2\ell$, each such block is 
the identity matrix; however, the block corresponding to a 
physical mode $0\leq q \leq 2\ell$ has the form
\begin{equation}
\left[
\begin{array}{c}
\mathbf{0}\\
\mathbf{0}\\
\hline
\mathsf{B}^{\ell q}
\end{array}
\right].
\end{equation}
Here $\mathbf{0}$ represents a row of zeros, and 
$\mathsf{B}^{\ell q}$ is a nonzero $(N_r-1)$-by-$(N_r+1)$
submatrix. 
The zeros in the first two rows are filled in with 
the Dirichlet boundary conditions on
$\partial O^-$, using $\delta^-$, and the radiation boundary conditions
on $\partial O^+$, using  (\ref{eq:rmaxuBCs}). Since these radiation conditions 
mix one cosine ($q= 2m-1$) and the other 
sine ($q=2m$) mode, the tau conditions lead to a coupling between 
among the blocks. The resulting $2(N_r+1)$-by-$2(N_r+1)$
block neighborhood, with Dirichlet and radiation boundary 
conditions, takes one of the following forms
(either representation is possible
due to the homogeneity of the boundary conditions):
\begin{equation}
\left[
\begin{array}{c|c}
\begin{array}{c}
\delta^-\\
p\delta^+\\
\mathsf{B}^{\ell,2m-1}
\end{array}
&
\begin{array}{c}
\mathbf{0}\\
R\nu^+ + q\delta^+\\
\mathbf{0}
\end{array}
\\
\hline
\begin{array}{c}
\mathbf{0}\\
R\nu^+ + q\delta^+\\
\mathbf{0}
\end{array}
&
\begin{array}{c}
\delta^-\\
-p\delta^+\\
\mathsf{B}^{\ell,2m}
\end{array}
\end{array}
\right]
\quad \mbox{or}\quad
\left[
\begin{array}{c|c}
\begin{array}{c}
\delta^-\\
R\nu^+ + q\delta^+\\
\mathsf{B}^{\ell,2m-1}
\end{array}
&
\begin{array}{c}
\mathbf{0}\\
-p\delta^-\\
\mathbf{0}
\end{array}
\\
\hline
\begin{array}{c}
\mathbf{0}\\
p\delta^+\\
\mathbf{0}
\end{array}
&
\begin{array}{c}
\delta^-\\
R\nu^+ + q\delta^+\\
\mathsf{B}^{\ell,2m}
\end{array}
\end{array}
\right],
\label{Ablock}
\end{equation}
where $\mathbf{0}$ represents either a row  
(when opposite a $\delta^{-}$) or 
a $(N_r-1)$-by-$(N_r+1)$ submatrix of zeros 
(when opposite a ${\mathsf B}$). 
 Boundary
conditions for $m=0$ (zero modes) correspond to blocks
\begin{equation}
\left[
\begin{array}{c}
\delta^-\\
R\nu^+ + q\delta^+\\
\hline
\mathsf{B}^{\ell 0}
\end{array}
\right].
\end{equation}
Evidently, in this case only a single azimuthal block need be considered.

\subsection{Inner spherical shells}\label{subsec:innershells}
We stress that the polar coordinates $(r,\theta,\phi)$ appearing in
this subsection are {\em not} the polar coordinates $(r,\theta,\varphi)$ 
used in the last, although only the notation for the azimuthal angle 
($\phi$ vs.~$\varphi$) reflects the difference. We start with 
\eqref{eqn:HRWEoperator}, assume 
that one of the ``holes'' is at $\widetilde{Z} = z_H$, and define new
comoving 
coordinates
\begin{equation}
     z = \widetilde{Z}-z_H,\quad
     x = \widetilde{X},\quad
     y = \widetilde{Y}\,.
\end{equation}
The
helically reduced wave operator in the new coordinates is
\begin{equation}
L = \frac{\partial^2}{\partial x^2}
   +\frac{\partial^2}{\partial y^2}
   +\frac{\partial^2}{\partial z^2}
   -\Omega^2\left[
       (z_H+z)\frac{\partial}{\partial x}
      - x\frac{\partial}{\partial z}
           \right]^2.
\label{hrwo}
\end{equation}
Spherical polar coordinates $\{r,\theta,\phi\}$ in this subsection
correspond to the system $\{x,y,z\}$.

As already mentioned, the system $\{r,\theta,\phi\}$
is not the system $\{r,\theta,\varphi\}$ corresponding the outer
shell. Nevertheless, for an inner shell our treatment of the Laplacian 
part of the operator is the same as the treatment given in the last 
subsection. In particular, we adopt the same conventions for the
indexing of the spectral representation, and therefore again
arrive at the expression (\ref{eq:precondLapO}). Notationally, the 
only difference is that we replace all instances of $\varphi$ 
with $\phi$. Therefore, having already considered ($r^2\times$) 
the Laplacian part of the HRWE, we turn to ($r^2\times$) the term 
in (\ref{hrwo}) 
paired with $-\Omega^2$,
\begin{equation}\label{eq:JH}
J_H \equiv r^2[(z_H+z)\partial/\partial x
      - x\partial/\partial z]^2.
\end{equation} 

To facilitate the expression of the  derivatives in (\ref{eq:JH}) 
in terms of operations on $\{r,\theta,\phi\}$, we
introduce
\begin{align}
Q & = \sin\theta\cos\phi
\\
P & = \cos\theta\cos\phi\partial/\partial\theta 
- \csc\theta\sin\phi
\partial/\partial\phi\\
N & = \cos\phi\partial/\partial\theta - 
\cos\theta\csc\theta  \sin\phi\partial/\partial\phi,
\end{align}
and note that
\begin{equation}
\partial /\partial x = Q\partial /\partial r
+ r^{-1} P,\quad  
z\partial/\partial x
- x\partial/\partial z=N\,.
\end{equation}
With the identities (which should be read as operators 
acting on a scalar function) 
\begin{equation}
r^2 
\partial_r^2 
= \partial_r^2 r^2 - 4\partial_r r + 2, \quad
r\partial_r = \partial_r r 
-1,\quad
r^2 \partial_r = \partial_r r^2 - 2r,
\end{equation} 
we then find that $J_H$ of (\ref{eq:JH}) can be written as
\begin{align}
J_H 
& = z_H^2 Q^2 \partial_r^2 r^2
+ z_H^2(PQ+QP-4Q^2)\partial_r r + z_H (NQ+QN) \partial_r r^2
\nonumber \\
& + z_H (NP+PN - 2NQ - 2QN)r + N^2 r^2
\label{eq:JH2} \\
& + z_H^2 (P^2 + 2Q^2 - PQ - 2 QP).
\nonumber
\end{align}
We use $\mathcal{J}_H$ to denote the spectral form of the 
differential operator $J_H$. The corresponding sparse
form $\mathcal{B}\mathcal{J}_H \equiv (I_\theta \otimes I_\phi
\otimes B^2_{r[2]})\mathcal{J}_H$ is then
\begin{align}
\mathcal{B}\mathcal{J}_H
& = z_H^2 \mathsf{Q}^2 \otimes I_{[2]r} A_r^2
+ z_H^2(\mathsf{P}\mathsf{Q}+\mathsf{Q}\mathsf{P}-4\mathsf{Q}^2)
    \otimes B_{[2]r} A_r
\\
& + z_H (\mathsf{N}\mathsf{Q}
    +\mathsf{Q}\mathsf{N})\otimes B_{[2]r} A_r^2
+ z_H (\mathsf{N}\mathsf{P}+\mathsf{P}\mathsf{N}
  - 2(\mathsf{N}\mathsf{Q}
  +\mathsf{Q}\mathsf{N}))\otimes B^2_{[2]r} A_r
\nonumber \\
& + \mathsf{N}^2 \otimes B^2_{[2]r} A_r^2
+ z_H^2 (\mathsf{P}-2\mathsf{Q})(\mathsf{P}-\mathsf{Q})\otimes
B^2_{[2]r}.
\nonumber
\end{align}
Here the san serif $\mathsf{N}$,  $\mathsf{P}$, and
$\mathsf{Q}$ are matrices acting 
on the spectral space of spherical harmonic expansion 
coefficients. 
Whence we need explicit realizations of the following 
matrices: $\mathsf{Q}^2$,
$\mathsf{PQ}+\mathsf{QP}$,
$\mathsf{PN}+\mathsf{NP}$,
$\mathsf{PQ}+\mathsf{QP}$,
and $(\mathsf{P}-\mathsf{Q})^2$.
We compute these matrices as truncations of the corresponding 
exact infinite dimensional matrices described below (with 
products computed before truncation). 
The truncated matrix components 
$\mathsf{N}(\alpha,\beta)$ of $\mathsf{N}$
obey the following condition:
\begin{equation}
\mathsf{N}\big(\ell(2N_\theta+1)+ q,k(2N_\theta+1)+ p\big)
= 0,\qquad \text{ for } q > 2\ell \text{ or } p > 2k,
\end{equation}
and similarly for the components
$\mathsf{P}(\alpha,\beta)$ and 
$\mathsf{Q}(\alpha,\beta)$. (Here we have
switched to parenthesis notation \cite{GolubVanLoan} for the 
components $\mathsf{N}(i,j)
= \mathsf{N}_{ij}$ of a matrix.) This condition properly treats 
the extraneous components we have included in our expansion 
vector $\widetilde{\boldsymbol{\psi}}$.

Of the three angular differential operators $P,Q,N$, 
we only consider $N$ in detail here, as its action on spherical 
harmonics is the simplest to describe. Partial formulas are
given for $P$ and $Q$ at the end of this subsection. Using 
standard formulas from the theory of angular momentum
(see the appendix of 
\cite{HL}), we have
\begin{equation}\label{eq:Naction}
N Y_{\ell m}
= \half\sqrt{(\ell-m)(\ell+m+1)}Y_{\ell,m+1} -
  \half\sqrt{(\ell+m)(\ell-m+1)}Y_{\ell,m-1}.
\end{equation}

Before completing our construction of
$\mathsf{N}$, $\mathsf{P}$, and
$\mathsf{Q}$, we first collect some formulas which relate the 
standard complex representation of spherical harmonics
$Y_{\ell m}(\theta,\phi)$ to the real-valued
representation. The normalized Legendre functions are \cite{AS}
\begin{equation}
\overline{P}_{\ell m}(u) = (-1)^m\sqrt{\frac{2\ell+1}{2}
\frac{(\ell-m)!}{(\ell+m)!}} P^m_\ell(u),
\end{equation}
with $P^m_\ell(u)$ the standard associated Legendre function
(as given, for example, by Thorne \cite{Thorne}). We then have
\begin{equation}
Y_{\ell m} = \sqrt{\frac{1}{2\pi}}(-1)^m \overline{P}_{\ell m}
e^{\mathrm{i}m\phi}, \quad
Y_{\ell, -m} = \sqrt{\frac{1}{2\pi}}
\overline{P}_{\ell m} e^{-\mathrm{i}m\phi},\quad
m \geq 0.
\end{equation}

For fixed $\ell$,  the expansion in azimuthal index takes the form
\begin{align}
c_{\ell 0} Y_{\ell 0} & + \sum_{m=1}^\ell
\big(c_{\ell m} Y_{\ell m} + c_{\ell,-m} Y_{\ell,-m}\big)
= a_{\ell 0} \overline{P}_{\ell 0} +
\sum_{m=1}^\ell \overline{P}_{\ell m}
\big[a_{\ell m}\cos m\phi
+ b_{\ell m}\sin m\phi\big]\,,
\end{align}
where the real expansion coefficients are
$\sqrt{2\pi}a_{\ell 0} = c_{\ell0}$ and, for $m \geq 1$,
\begin{equation}
\sqrt{2\pi}a_{\ell m} = 
c_{\ell m}(-1)^m + c_{\ell,-m},\qquad
\sqrt{2\pi}b_{\ell m} = 
\mathrm{i}\big[c_{\ell m}(-1)^m-c_{\ell,-m}\big].
\end{equation}
We define another set of complex expansion coefficients
\begin{equation}
f_{\ell m} =
\half\sqrt{(\ell+m)(\ell-m+1)}c_{\ell,m-1} -
\half\sqrt{(\ell-m)(\ell+m+1)}c_{\ell,m+1}\,,
\end{equation}
so that, from (\ref{eq:Naction}), the action of $N$ has the effect
\begin{equation}
\Psi = \sum_{\ell = 0}^\infty \sum_{m=-\ell}^\ell
c_{\ell m} Y_{\ell m},\quad 
N\Psi = \sum_{\ell = 0}^\infty 
\sum_{m=-\ell}^\ell
f_{\ell m} Y_{\ell m}.
\end{equation}
We can then represent 
 $N$ by  the matrix $\mathsf{N}$ that converts the vector of coefficients
$c_{\ell m}$ to the vector $f_{\ell m}$ by $\boldsymbol{f} = 
\mathsf{N}\boldsymbol{c}$. We also define real coefficients
$d_{\ell m}$, $e_{\ell m}$ which are related to $f_{\ell m}$ in 
the same way that $a_{\ell m}$, $b_{\ell m}$ are related to 
$c_{\ell m}$, and then view the action of $\mathsf{N}$ as a 
mapping from the real coefficients $a_{\ell m}, b_{\ell m}$ to the 
real coefficients $d_{\ell m}, e_{\ell m}$. 

Turning to the representation for $Q$, we likewise use results 
tabulated in the appendix of \cite{HL} to find
\begin{align}
Q Y_{\ell m} & =
  \half \sqrt{\frac{(\ell-m+1)(\ell-m+2)}{(2\ell+1)(2\ell+3)}} Y_{\ell+1,m-1} \\
& - \half \sqrt{\frac{(\ell+m+1)(\ell+m+2)}{(2\ell+1)(2\ell+3)}}Y_{\ell+1,m+1}\nonumber 
\\ & - \half \sqrt{\frac{(\ell+m)(\ell+m-1)}{(2\ell+1)(2\ell-1)}}Y_{\ell-1,m-1} 
\nonumber \\
& + \half \sqrt{\frac{(\ell-m)(\ell-m-1)}{(2\ell+1)(2\ell-1)}} Y_{\ell-1,m+1}, 
\nonumber
\end{align}
so that $\boldsymbol{f} = \mathsf{Q}\boldsymbol{c}$ is determined by
\begin{align}
f_{\ell m} & =
    \half \sqrt{\frac{(\ell-m-1)(\ell-m)}{(2\ell-1)(2\ell+1)}} c_{\ell-1,m+1} \\
& - \half \sqrt{\frac{(\ell+m-1)(\ell+m)}{(2\ell-1)(2\ell+1)}} c_{\ell-1,m-1}\nonumber\\ 
& - \half \sqrt{\frac{(\ell+m+2)(\ell+m+1)}{(2\ell+1)(2\ell+3)}} c_{\ell+1,m+1}\nonumber \\
& + \half \sqrt{\frac{(\ell-m+2)(\ell-m+1)}{(2\ell+1)(2\ell+3)}} c_{\ell+1,m-1}.
\nonumber
\end{align}
Again, we may express the action of $\mathsf{Q}$ as a mapping from
$a_{\ell m}$, $b_{\ell m}$ to $d_{\ell m}$, $e_{\ell m}$.
Finally, we use the identity
\begin{align}
P Y_{\ell m} =  &
-\half \ell\sqrt{\frac{(\ell-m+1)(\ell-m+2)}{(2\ell+1)(2\ell+3)}} 
Y_{\ell+1,m-1} \\
& + \half \ell \sqrt{\frac{(\ell+m+1)(\ell+m+2)}{(2\ell+1)(2\ell+3)}}
    Y_{\ell+1,m+1}\nonumber\\ 
& - \half (\ell+1) 
\sqrt{\frac{(\ell+m)(\ell+m-1)}{(2\ell+1)(2\ell-1)}}Y_{\ell-1,m-1}
\nonumber \\
& + \half (\ell+1) \sqrt{\frac{(\ell-m)(\ell-m-1)}{(2\ell+1)(2\ell-1)}} 
Y_{\ell-1,m+1}.
\nonumber
\end{align}
to similarly define the action of 
$\mathsf{P}$ as a mapping from $a_{\ell m}$, $b_{\ell m}$ to
$d_{\ell m}$, $e_{\ell m}$.

To enforce the inner and outer boundary conditions in
(\ref{eq:introHRWE}), we fill empty rows in 
$(\mathcal{B}\mathcal{A}_r^2 \Delta)^\mathrm{modified}
-\Omega^2 \mathcal{B}\mathcal{J}_H$ with the tau-conditions.
Let $h^{+}(\theta,\phi) = 
\psi(r_\mathrm{max},\theta,\phi)$ and 
$h^{-}(\theta,\phi) =
\psi(r_\mathrm{min},\theta,\phi)$. Then Dirichlet
boundary conditions on the inner and outer boundaries
of the shell are expressible as
\begin{equation}\label{eq:innershellBCs}
\sum_{n=0}^{N_r} \widetilde{\psi}_{\ell q n} \delta_n^{\pm}
= \tilde{h}^\pm_{\ell q},
\end{equation}
where spherical-harmonic projections appear on the righthand
side. Table \ref{tab:shellrows} shows how empty rows are filled
to enforce these boundary conditions.
\begin{table}
\begin{tabular}{|c|c|c|}
\hline
Boundary & Rows & Index restrictions \\
\hline
$r = r_\mathrm{min}$       & $\ell(2N_\theta+1)(N_r+1)+q(N_r+1)+0$ & $0\leq \ell\leq N_\theta,\; 0\leq q\leq 2\ell$ \\
$r = r_\mathrm{max}$       & $\ell(2N_\theta+1)(N_r+1)+q(N_r+1)+1$ & $0\leq \ell\leq N_\theta,\; 0\leq q\leq 2\ell$ \\
\hline
\end{tabular}
\caption{{\sc Filling of empty rows for shells.}
\label{tab:shellrows}
}
\end{table}

\subsection{Cylindrical shells}\label{subsec:cylinders}
Throughout this section we suppress the tildes on
$\widetilde{X},\widetilde{Y},\widetilde{Z}$. Let $\rho 
= (X^2 + Y^2)^{1/2}$, and multiply 
Eq.~\eqref{eqn:HRWEoperator} by $\rho^2$ to get the
operator identity
\begin{equation}\label{eqn:rho2HRWE}
\rho^2L = \rho^2\big[\partial^2_Y 
+ \partial^2_X + \partial^2_Z
- \Omega^2 \partial^2_X Z^2
- \Omega^2 \partial^2_Z X^2
- \Omega^2 (\partial_X X
          +\partial_Z Z
         -2\partial_X X 
           \cdot 
           \partial_Z Z)
\big].
\end{equation}
Since $X = \rho\cos\phi$ and $Y = \rho\sin\phi$,
\begin{equation}\label{eqn:LaplPartC}
\rho^2(\partial^2_X+\partial^2_Y +\partial_Z^2) = \partial^2_\rho\rho^2
- 3\partial_\rho \rho + 1 + \partial^2_\phi + \rho^2\partial_Z^2,
\end{equation}
again with the view that this is an operator identity. 
Eq.~\eqref{eqn:LaplPartC} has the spectral representation 
\begin{equation}\label{eqn:MatrixLaplPartC}
\mathcal{A}_\rho^2\boldsymbol{\Delta}
= I_\phi \otimes (I_\rho + D^2_\rho A_\rho^2
- 3D_\rho A_\rho) \otimes I_Z
+ D^2_\phi \otimes I_\rho \otimes I_Z
+ I_\phi \otimes A_\rho^2 \otimes D_Z^2.
\end{equation}
Here $\mathcal{A}_\rho^2$ represents the matrix 
$I_\phi \otimes A_\rho^2 \otimes I_Z$.
To achieve this representation we start by 
introducing a mapping of 
$[Z_\mathrm{min},Z_\mathrm{max}]$ to $[-1,1]$ with the function
$\chi(Z)$, so that $Z$ dependence can be expressed with the
Chebyshev polynomials $T_p(\chi(Z))$. Similarly $\xi(\rho)$ 
maps $[\rho_\mathrm{min},\rho_\mathrm{max}]$ to $[-1,1]$. The 
solution is then formally expressed as
\begin{align}
\psi(\rho,\phi,Z) & =
\sum_{n=0}^\infty\sum_{p=0}^\infty 
\widetilde{\psi}_{0np}T_n(\xi(\rho))T_p(\chi(Z))
\nonumber \\
& +
\sum_{k=1}^{\infty} \sum_{n=0}^\infty \sum_{p=0}^\infty
\big[\widetilde{\psi}_{2k-1,np}\cos(k\theta)
    +\widetilde{\psi}_{2k,np}\sin(k\theta)\big]
T_n(\xi(\rho))T_p(\chi(Z)),
\end{align}
with corresponding numerical truncation (taking $N_\phi$ even for
simplicity)
\begin{align}
\mathcal{P}_{N_\rho,N_\phi,N_z}\psi(\rho,\phi,Z) & =
\sum_{n=0}^{N_\rho}\sum_{p=0}^{N_z}
\widetilde{\psi}_{0np}T_n(\xi(\rho))T_p(\chi(Z))
\nonumber \\
& +
\sum_{k=1}^{\frac{1}{2}N_\phi} \sum_{n=0}^{N_\rho} \sum_{p=0}^{N_z}
\big[\widetilde{\psi}_{2k-1,np}\cos(k\phi)
    +\widetilde{\psi}_{2k,np}\sin(k\phi)\big]
T_n(\xi(\rho))T_p(\chi(Z)).
\label{eqn:cyltrunc}
\end{align}
The direct product structure in
\eqref{eqn:MatrixLaplPartC} has been determined by the convention
\begin{equation}\label{eq:dpFORcylinders}
\widetilde{\boldsymbol{\psi}}(m(N_\rho+1)(N_z+1) + n(N_z+1) + p) 
= \widetilde{\psi}_{mnp}.
\end{equation}
We now take the matrix representation of the differential operator
in (\ref{eqn:rho2HRWE}), and ``sparsify'' it via multiplication by
$\mathcal{B} = B^2_\phi \otimes B^2_{\rho[2]} \otimes B^2_{Z[2]}$. In this 
operator the $B^2_{\rho[2]}$ and $B^2_{Z[2]}$ correspond to the 
usual $B$ operators representing integration twice over a coordinate, 
and leaving two empty rows to be filled by tau-conditions. 
The operator $B^2_\phi$ applies to a coordinate with no endpoints, and 
hence no applicable boundary conditions. It represents double integration
over all Fourier modes except the zero mode, which is left unchanged. The 
matrix that accomplishes this
has the explicit form
$B^2_\phi = \mathrm{diag}(1,-1,-1,
-\frac{1}{4},-\frac{1}{4},-\frac{1}{9},-\frac{1}{9},\cdots)$. 
Although the operation on $\phi$ does not play a role in handling 
boundary conditions, it should further enhance the spectrum 
clustering of the matrix that must be inverted.

With this $\mathcal{B}$ the sparsified form of (\ref{eqn:MatrixLaplPartC})
becomes
\begin{equation}\label{eqn:BMatrixLaplPartC}
\mathcal{B} \mathcal{A}_\rho^2\boldsymbol{\Delta}
= B^2_\phi \otimes (B^2_{\rho[2]} + I_{\rho[2]} A_\rho^2
- 3 B_{\rho[2]} A_\rho) \otimes B^2_{Z[2]}
+ I_{\phi[1]} \otimes B^2_{\rho[2]} \otimes B^2_{Z[2]}
+ B^2_\phi \otimes B^2_{\rho[2]} A_\rho^2 \otimes I_{Z[2]}.
\end{equation}
Our analysis of the terms in the HRWE 
proportional to $\Omega^2$ starts with the expressions
\begin{align}
\rho^2\partial_X &= \rho^2\cos\phi\partial_\rho - \rho\sin\phi\partial_\phi
\\
\rho^2 \partial^2_X
 & = \rho^2\cos^2\phi \partial_\rho^2
-2\rho\cos\phi\sin\phi \partial_\rho\partial_\phi 
+ 2\cos\phi\sin\phi \partial_\phi
+ \rho\sin^2\phi\partial_\rho + \sin^2\phi \partial_\phi^2.
\end{align}
With the operator identities $\rho^2 \partial_\rho^2
= \partial_\rho^2 \rho^2 - 4\partial_\rho \rho + 2$,
$\rho\partial_\rho = \partial_\rho \rho -1$, and
$\rho^2 \partial_\rho = \partial_\rho \rho^2 - 2\rho$, we then find
\begin{align}
\rho^2\partial_X & = \partial_\rho \rho^2\cos\phi 
                   - \rho (2\cos\phi + \sin\phi\partial_\phi)
\label{eqn:rho2d_X}
\\
\rho^2 \partial^2_X
& = \partial_\rho^2\rho^2\cos^2\phi
+   \partial_\rho \rho (1- 5\cos^2\phi 
-   2\cos\phi\sin\phi\partial_\phi)
\nonumber \\
& + \sin^2\phi\partial_\phi^2
  + 4\cos\phi\sin\phi \partial_\phi
  + 3\cos^2\phi - 1.
\label{eqn:rho2d_XX}
\end{align}
Still viewing these relationships as operator identities, we now
exploit the product rule for differentiation to rewrite the terms
involving angular derivatives, with the results
\begin{align}
\rho^2\partial_X & = 
\partial_\rho \rho^2\cos\phi
- \rho (\cos\phi + \partial_\phi\sin\phi)
\label{eqn:rho2d_Xnew}
\\
\rho^2 \partial^2_X
& = \partial_\rho^2\rho^2\cos^2\phi
-   \partial_\rho \rho (1 + \cos^2\phi 
+ 2\partial_\phi\cos\phi\sin\phi)
+ \partial^2_\phi \sin^2\phi + \sin^2\phi.
\label{eqn:rho2d_XXnew}\end{align}
These equations then yield
\begin{subequations}\label{eqn:rho2d_XonX}
\begin{align}
\rho^2\partial_X X & = \partial_\rho \rho^3\cos^2\phi
                   - \rho^2 (\cos^2\phi  
                   + \partial_\phi\cos\phi\sin\phi) 
\\
\rho^2 \partial^2_X
& = (\partial_\rho^2\rho^2 - 3\partial_\rho\rho + 1)\cos^2\phi
\nonumber \\
& -   \partial_\rho \rho (1 - 2\cos^2\phi
+ 2\partial_\phi\cos\phi\sin\phi)
+ \partial^2_\phi \sin^2\phi + 2\sin^2\phi - 1,
\end{align}
\end{subequations}
expressions which prove useful in obtaining the matrix 
representation of the operator on the righthand side 
of \eqref{eqn:rho2HRWE}. 
To optimize the implementation, we have chosen the first 
term on the righthand side of (\ref{eqn:rho2d_XonX}b)
to match a similar term in the Laplacian part of
the operator [cf.~Eq.~\eqref{eqn:LaplPartC}].

We split the HRWE operator on a cylindrical shell as $\rho^2L = 
\rho^2\Delta - \Omega^2 J$, where
\begin{equation}
J = J_1 + J_2 + J_3 + J_4 =
    \underbrace{\rho^2 \partial_X X (1 - 2 \partial_Z Z)}_{J_1} +
    \underbrace{\rho^2\partial^2_X Z^2}_{J_2} + 
    \underbrace{\rho^2 \partial_Z Z}_{J_3} +
    \underbrace{\rho^2\partial^2_Z X^2}_{J_4}.
\end{equation}
The piece $\rho^2\Delta$ was shown to lead to (\ref{eqn:BMatrixLaplPartC}). 
We now focus on $J$ whose matrix representation stems from 
the representations of its constituents. Equations~\eqref{eqn:rho2d_XXnew} 
and \eqref{eqn:rho2d_XonX} give us
\begin{align}
\mathsf{J}_1 
& = C^2_\phi \otimes
     D_\rho A_\rho^3 \otimes
     ( I_Z - 2 D_Z A_Z )
   - (C^2_\phi + D_\phi C_\phi S_\phi) \otimes
     A_\rho^2 \otimes
     (I_Z - 2 D_Z A_Z ) \\
\mathsf{J}_2
& = C^2_\phi \otimes (I + 
    D_\rho^2 A_\rho^2 
- 3 D_\rho A_\rho) \otimes 
    A_Z^2
-   (I_\phi - 2C^2_\phi
+   2D_\phi C_\phi S_\phi) \otimes 
    D_\rho A_\rho \otimes A_Z^2
\\
& + (D_\phi^2S^2_\phi + 2S^2_\phi - 1) \otimes 
    I_\rho \otimes 
    A_Z^2 
\nonumber \\
\mathsf{J}_3 & = I_\phi \otimes A_\rho^2 \otimes D_Z A_Z \\
\mathsf{J}_4 & = C^2_\phi \otimes A_\rho^4 \otimes D^2_Z,
\end{align}
where $S_\phi$ and $C_\phi$ are respectively the matrices in the
Fourier basis which correspond to multiplication by $\sin\phi$
and $\cos\phi$. Applying the sparsifying matrix 
$\mathcal{B} = B^2_\phi \otimes B^2_{\rho[2]} \otimes B^2_{Z[2]}$, 
we then have
\begin{align}
\mathcal{B}\mathsf{J}_1
& =  B^2_\phi C^2_\phi \otimes
     B_{\rho[2]} A_\rho^3 \otimes
     ( B^2_{Z[2]} - 2 B_{Z[2]} A_Z )
\\
& - (B^2_\phi C^2_\phi + B_{\phi [1]} C_\phi S_\phi) 
     \otimes
     B^2_{\rho [2]} A_\rho^2 \otimes
     (B^2_{Z[2]} - 2 B_{Z[2]} A_Z ) 
\nonumber \\
\mathcal{B}\mathsf{J}_2
& = B^2_\phi C^2_\phi \otimes
    (B^2_{r[2]} + I_{\rho [2]} A_\rho^2 
                - 3B_{r[2]}A_\rho )\otimes
    B^2_{Z[2]} A_Z^2
\\
& - (B^2_\phi - 2 B^2_\phi C^2_\phi
  + 2 B_{\phi [1]} C_\phi S_\phi) \otimes
    B_{\rho [2]} A_\rho \otimes B^2_{Z [2]} A_Z^2
\nonumber \\
& + (I_{\phi [1]} S^2_\phi
  + 2 B^2_\phi S^2_\phi - B^2_\phi) \otimes
    B^2_{\rho [2]} \otimes
    B^2_{Z [2]}  A_Z^2 
\nonumber \\
\mathcal{B}\mathsf{J}_3 & = 
    B^2_\phi \otimes 
    B^2_{\rho [2]} A_\rho^2 \otimes 
    B_{Z [2]} A_Z \\
\mathcal{B}\mathsf{J}_4 & = 
    B^2_\phi C^2_\phi \otimes 
    B^2_{\rho [2]} A_\rho^4 \otimes 
    I_{Z[2]}.
\end{align}
The sparsified matrix representing  (\ref{eqn:rho2HRWE}) is then    
$\mathcal{B}\mathcal{A}_\rho^2 \mathcal{L} = 
\mathcal{B}\mathcal{A}_\rho \boldsymbol{\Delta} - \Omega^2 
(\mathcal{B}\mathsf{J}_1
+\mathcal{B}\mathsf{J}_2
+\mathcal{B}\mathsf{J}_3
+\mathcal{B}\mathsf{J}_4)$.
\begin{table}
\begin{tabular}{|l|l|l|}
\hline
\multicolumn{1}{|c|}{Boundary} & 
\multicolumn{1}{|c|}{Rows} & 
\multicolumn{1}{|c|}{Index restrictions} \\
\hline
$\rho = \rho_\mathrm{min}$ & $m(N_\rho+1)(N_z+1) + p$            & $0\leq m\leq N_\phi,\; 2\leq p\leq N_z$\\
$\rho = \rho_\mathrm{max}$ & $m(N_\rho+1)(N_z+1) + (N_z+1) + p$  & $0\leq m\leq N_\phi,\; 2\leq p\leq N_z$ \\
\hline
$Z = Z_\mathrm{min}$       & $m(N_\rho+1)(N_z+1) + n(N_z+1) + 0$ & $0\leq n\leq N_\rho,\; 0\leq m\leq N_\phi$ \\
$Z = Z_\mathrm{max}$       & $m(N_\rho+1)(N_z+1) + n(N_z+1) + 1$ & $0\leq n\leq N_\rho,\; 0\leq m\leq N_\phi$ \\
\hline
\end{tabular}
\caption{{\sc Filling of empty rows for cylinders.}
\label{tab:cylinderrows}
}
\end{table}

To enforce boundary conditions, we fill empty rows in
the matrix $\mathcal{B}\mathcal{A}_\rho \boldsymbol{\Delta} - 
\Omega^2\mathcal{B}\mathsf{J}$ with the
tau-conditions. Let $h^{+}(\phi,Z) =
\psi(\rho_\mathrm{max},\phi,Z)$,
$h^{-}(\phi,Z) = \psi(\rho_\mathrm{min},\phi,Z)$
and $f^{+}(\rho,\phi) =
\psi(\rho,\phi,Z_\mathrm{max})$,
$f^{-}(\rho,\phi,Z_\mathrm{min}) 
= \psi(\rho,\phi,Z_\mathrm{min})$.
Then Dirichlet boundary conditions on the
inner and outer axial boundaries and on the top and
bottom caps are expressible as
\begin{equation}\label{eq:cylinderBCs}
\sum_{n=0}^{N_\rho} \widetilde{\psi}_{mnp} \delta_n^{\pm}
= \tilde{h}^\pm_{mp},\qquad 
\sum_{p=0}^{N_z} \widetilde{\psi}_{mnp} \delta_p^{\pm}
= \tilde{f}^\pm_{mn}.
\end{equation}
There are $(N_\phi+1)(N_z+1) + (N_\rho+1)(N_\phi+1)$ such
equations possible. However, owing to the fact that the
caps shares common edges with both the inner and outer 
axial boundaries, there are $2(N_\phi+1)$ linear dependencies
amongst these equations, and in fact the number of available
empty rows is precisely
$$
(N_\phi+1)(N_Z+1) + (N_\rho+1)(N_\phi+1) - 2(N_\phi+1).
$$
Table \ref{tab:cylinderrows} shows how we fill zero rows to
enforce the boundary conditions.

\section{Gluing of subdomains}\label{sec:glue}
So far we have described individual shell, cylinder, and block
subdomains (and their associated tau-conditions) as if they were
decoupled. All the subdomains are, of course, coupled and we refer 
to the process of making them parts of a single problem as ``gluing.''
Matching, or gluing,  must be done for each subset of subdomains 
that touch, whether that touching is a finite volume overlap or 
a lower-dimensional shared boundary. The global problem requires 
matching for the following subdomain configurations:
\begin{itemize}
\item[(i)]  Two adjacent cylinders. 
\item[(ii)] One inner shell and a combination of
             cylinders and blocks.
\item[(iii)] One cylinder and one block.
\item[(iv)] The outer shell and the combination of 
            blocks $B$ and $D$ and all cylinders.
\end{itemize}
We describe (i) and (ii) in detail, provide a sketch of 
(iii), and omit a description of (iv) altogether. Although more 
complicated, a description of (iv) would parallel that of (iii). 

Before giving more details, we comment on how such 
gluing is reflected in the overall linear system. Let, for 
example, $\widetilde{\boldsymbol{\psi}}{}^J$ and 
$\widetilde{\boldsymbol{\psi}}{}^B$ respectively represent 
the vectors of spherical-harmonic Chebyshev and triple 
Chebyshev expansion coefficients corresponding to the inner 
shell $J$ and block $B$ of Fig.~\ref{fig:domaindecomp3}. 
The overall set of unknowns is the concatenation 
$$\widetilde{\mathbf{\Psi}}=
(\widetilde{\boldsymbol{\psi}}{}^J,
\widetilde{\boldsymbol{\psi}}{}^H,
\widetilde{\boldsymbol{\psi}}{}^B,
\widetilde{\boldsymbol{\psi}}{}^C,
\widetilde{\boldsymbol{\psi}}{}^D,
\widetilde{\boldsymbol{\psi}}{}^1,
\widetilde{\boldsymbol{\psi}}{}^2,
\widetilde{\boldsymbol{\psi}}{}^3,
\widetilde{\boldsymbol{\psi}}{}^4,
\widetilde{\boldsymbol{\psi}}{}^5,
\widetilde{\boldsymbol{\psi}}{}^O)^t,
$$ 
which satisfies a linear system stemming from 
Eq.~(\ref{eq:introHRWE}),
\begin{equation}\label{eq:2centerLinearSystem}
\mathcal{M}\widetilde{\mathbf{\Psi}} = 
\mathcal{B}\widetilde{\mathcal{G}},
\end{equation}
where $\widetilde{\mathcal{G}}$ is a similar concatenation of 
the sources $\widetilde{\mathbf{g}}$ on the individual subdomains. 
Here $\mathcal{B}$ indicates integration ``preconditioning"
(sparsification) on all subdomains. Symbolically then, 
the coefficient matrix $\mathcal{M}$ 
is $\mathcal{B}\mathcal{L}$, now with $\mathcal{L}$ standing for the 
spectral representation of the HRWE operator $L$ on the whole 
2-center domain. In this symbolic view, we have ignored multiplications 
by radial powers on spheres and cylinders.

Each of the eleven subdomains in Fig.~\ref{fig:domaindecomp3} is 
represented by one of eleven super-blocks ($J$-$J$, $H$-$H$,
$\cdots$, $O$-$O$) which sit along the 
diagonal of the overall super-matrix $\mathcal{M}$ representing the 
PDE on the whole 2-center domain. We use the term ``super-block'' 
here since the matrix corresponding to each subdomain arises, as we 
have seen, from a direct
product structure (and so could be viewed as already in a block
form). The supplementary equations needed for gluing are placed within
existing empty rows in the same manner as for the
tau-conditions. However, the gluing conditions stretch beyond the
super-block diagonal, since they are linear relationships between the
spectral expansion coefficients on two (or more) separate subdomains. For
example, the gluing together of cylinders 1 and 2 (which share a 
common cap) involves not only filling rows within the 
1-1 and 2-2 super-blocks along the diagonal of $\mathcal{M}$, but 
also filling rows within the 1-2 and 2-1 off-diagonal super-blocks.

\subsection{
Gluing of cylinders to cylinders}\label{subsec:glueCyls}
As a specific example, let us consider the gluing of 
cylinders 1 and 2 in Fig.~\ref{fig:domaindecomp3}, 
which as indicated share the cap $Z = Z_*$, where 
$Z_*$ is $Z_\mathrm{max}$ for cylinder 1 and 
$Z_\mathrm{min}$ for cylinder 2 (the common
cap has a hole in the middle, since 1 and 2 are 
cylindrical shells). Let, for example,
$\mathcal{P}\psi^1$ be shorthand for the numerical 
solution $\mathcal{P}_{N_\rho^1,N_\phi^1,N_Z^1}\psi^1$
for cylinder 1, as expressed in (\ref{eqn:cyltrunc}).
The restriction $\mathcal{P}\psi^1(\rho,\phi,Z_*)$
is a two-variable function on the cap $Z=Z_*$,
and it can be expanded in a {finite} 
Fourier-Chebyshev series, with $\sum_{k=0}^{N_Z^1}
\widetilde{\psi}{}^1_{qnk} \delta_k^+$ as the
corresponding two-index modal coefficients. Likewise,
the restriction $(d\mathcal{P}\psi^1/dZ)(\rho,\phi,Z_*)$
of the $Z$-derivative has a Fourier-Chebyshev series 
with two-index modal coefficients
$\sum_{k = 0}^{N_Z^1} \widetilde{\psi}{}^1_{qnk} \alpha_1\nu_k^+$.
The $\alpha_1$ factor is a scaling of the Neumann vector
$\nu^+$, and its presence is necessary since the range 
of $Z$ is not $[-1,1]$ (details are given in \cite{LauPriceI}).

On the $Z=Z_*$ cap we likewise consider the numerical solution
$\mathcal{P}\psi^2(\rho,\phi,Z_*)$ and its $Z$-derivative
$(d\mathcal{P}\psi^2/dZ)(\rho,\phi,Z_*)$, as determined by the
numerical solution $\mathcal{P}\psi^2$ on cylinder 2.  We distinguish
between two cases: (i) both the $N_\rho$ and $N_\phi$ truncations are
the same for cylinders 1 and 2 (but $N_Z^1 \neq N_Z^2$ is allowed),
and (ii) at least one of these truncations differs between the two
cylinders (i.e.~either $N_\rho^1 \neq N_\rho^2$ or $N_\phi^1 \neq
N_\phi^2$, or both, hold). Let us first consider case (i), returning
to case (ii) in the next paragraph.  For case (i) both
$\mathcal{P}\psi^1(\rho,\phi,Z_*)$ and
$\mathcal{P}\psi^2(\rho,\phi,Z_*)$ have two-surface modes 
which are in one-to-one correspondence, and likewise for 
the derivatives. Therefore, for this case we enforce\footnote{
We regret an error in the definition of $\nu^-$ in
Ref.~\cite{LauPriceI}, Eq.~(42). The correct expressions are
$
\nu^{\pm}=
\left[T_0^\prime(\pm1),
      T_1^\prime(\pm1),
      T_2^\prime(\pm1),
      T_3^\prime(\pm1),
      T_4^\prime(\pm1),\cdots\right]
=\left[0,1,\pm4,9,\pm16,\cdots\right].
$
In Ref.~\cite{LauPriceI} the righthand side of the second 
equation of (69) is also off by a sign.
}
\begin{equation}
\sum_{k = 0}^{N_Z^1}\widetilde{\psi}{}^1_{qnk} \delta_k^+ =
\sum_{k = 0}^{N_Z^2}  \widetilde{\psi}{}^2_{qnk} \delta_k^-,
\qquad
\sum_{k = 0}^{N_Z^1} \widetilde{\psi}{}^1_{qnk} \alpha_1\nu_k^+ =
\sum_{k = 0}^{N_Z^2} \widetilde{\psi}{}^2_{qnk} \alpha_2 \nu_k^-,
\label{1-2tau}
\end{equation}
for each Fourier-Chebyshev index pair $(q,n)$.
Here, for case (i), the
matching conditions enforce continuity between
the finite representations $\mathcal{P}\psi^1$ and 
$\mathcal{P}\psi^2$ across the cap,
and also  continuity between the finite representations 
$d\mathcal{P}\psi^1/dZ$ and 
$d\mathcal{P}\psi^2/dZ$. These matching conditions  are reflected 
in the overall matrix $\mathcal{M}$ as follows. As the super-block 
corresponding to each of the subdomains 1 and 2 has been 
sparsified in the described
fashion, each has a collection of empty rows which are also empty
throughout $\mathcal{M}$. In, say, the empty rows stretching
across the 1-1 and 1-2 super-blocks, we insert the first set of conditions
given in (\ref{1-2tau}). In the empty rows stretching
across the 2-2 and 2-1 super-blocks, we similarly place the Neumann
conditions, the second set of conditions given in (\ref{1-2tau}).
This filling of empty rows to achieve the required matching 
consists of relationships of modal coefficients with no reference to 
any ``sources''; they are  homogeneous equations.

To better understand the issues which will arise in matching
volume-overlapping subdomains, we now consider case (ii), the case in which
the cylinders 1 and 2 give rise to a disparate set of surface modes on
the $Z=Z_*$ cap. In this case, for example, we again have
$\sum_{k=0}^{N_Z^1} \widetilde{\psi}{}^1_{qnk} \delta_k^+$ as the
modal coefficients determining $\mathcal{P}\psi^1(\rho,\phi,Z_*)$, and
$\sum_{k=0}^{N_Z^2} \widetilde{\psi}{}^2_{qnk} \alpha_2 \nu_k^-$
as the modal coefficients determining 
$(d\mathcal{P}\psi^2/dZ)(\rho,\phi,Z_*)$. Now, however,  
\eqref{1-2tau} is not applicable. 
Instead, we now fix [cf.~the first equation in \eqref{eq:cylinderBCs}]
\begin{equation}\label{eq:nonconfcyl}
\sum_{k=0}^{N_z^1} \widetilde{\psi}{}^1_{qnk}
\delta_k^{+}
= \tilde{f}^+_{qn}
\qquad
\tilde{e}^-_{qn} =
\sum_{k=0}^{N_Z^2} \widetilde{\psi}{}^2_{qnk}\alpha_2\nu_k^-,
\end{equation}
where here $\tilde{f}^+_{qn}$ (for $0 \leq q \leq N_\phi^1$
and $0 \leq n \leq N_\rho^1$) and $\tilde{e}^-_{qn}$
(for $0 \leq q \leq N_\phi^2$
and $0 \leq n \leq N_\rho^2$) are
not to be viewed as inhomogeneities, rather as expressions built 
respectively with the modal coefficients for 
$\mathcal{P}\psi^2(\rho,\phi,Z_*)$ and 
$(d\mathcal{P}\psi^1/dZ)(\rho,\phi,Z_*)$. Note that,
as with Eqs.~\eqref{1-2tau}, these equations have the
form ``cylinder 1 coefficients = cylinder 2 coefficients".

Let us consider only $\tilde{f}^+_{qn}$, since similar comments 
apply to $\tilde{e}^-_{qn}$. 
First, we start with cylinder 1 and
define a Chebyshev-Lobatto/Fourier grid $\{(\rho_j,\phi_i):
0\leq j\leq N_\rho^1, 0\leq i\leq N_\phi^1\}$ on the $Z=Z_*$ 
cap of cylinder 1. The use of these points affords a double discrete 
Fourier-Chebyshev transform, through numerical quadrature, relating 
function values at the points and mode coefficients. (In practice, 
we have exploited the trigonometric form of the Chebyshev polynomials 
and have used the FFT to define both the Fourier and Chebyshev 
components of this transform.) The double discrete transform allows 
us to express the modal coefficients 
$\tilde{f}^+_{qn}$ in terms of the function values $f_{ij}^+$, 
at $\rho_i,\phi_j$, for $Z=Z_*$ on cylinder 1, in a form
\begin{equation}\label{eq:fqn2}
\tilde{f}^+_{qn} = \sum_{i=0}^{N_\rho^1} \sum_{j=0}^{N_\phi^1}
\mathcal{F}_{qn,ij} f_{ij}^+ .
\end{equation}
Next, at the nodal points $(\rho_j,\phi_i)$ of cylinder 1,
we evaluate $f^+_{ij}$ in terms of the expansion for the solution on 
cylinder 2, thereby finding
\begin{equation}\label{eq:fij2}
f^+_{ij} = \mathcal{P}\psi^2(\rho_j,\phi_i,Z_*) = 
\sum_{q=0}^{N^2_\phi}\sum_{n=0}^{N^2_\rho}
\mathcal{E}_{ij,qn} \sum_{k=0}^{N^2_Z} \widetilde{\psi}{}^2_{qnk}
\delta^-_k .
\end{equation}
Here, the values $\mathcal{E}_{ij,qn}$ arise from the evaluations
of the modal functions (Chebyshev and Fourier) of cylinder 2 at the
nodal points $(\rho_j,\phi_i)$ of cylinder 1.  
When the expressions for $f^+_{ij}$ from (\ref{eq:fij2}) are substituted 
in (\ref{eq:fqn2}), we get expressions for $\tilde{f}^+_{qn}$ in terms 
of the modal coefficients $\widetilde{\psi}{}^2_{qnk}$ representing 
the solution in cylinder 2. Finally, 
we substitute this $\tilde{f}^+_{qn}$ into
(\ref{eq:nonconfcyl}), which yields relationships between the
modal coefficients on cylinder 1 and cylinder 2 that express 
continuity of the solution across $Z=Z_*$. 

The linear relationships \eqref{eq:nonconfcyl} would likewise be inserted
into the overall coefficient matrix $\mathcal{M}$. Similar to before, the
righthand side of the first equation in \eqref{eq:nonconfcyl} would fill
empty rows stretching across the 1-2 super-block, with the $\delta^+$ 
vectors on the lefthand side filling empty rows across the 1-1 super-block.
The relationships expressed in the second equation in \eqref{eq:nonconfcyl}
would fill empty rows stretching across the 2-2 and 2-1 super-blocks
Finally, we note that the equations (\ref{eq:nonconfcyl}) reduce to
(\ref{1-2tau}) when $N_\phi^1 = N_\phi^2$ and $N_\rho^1 = N_\rho^2$.

\subsection{Gluing of an inner shell to cylinders and blocks} The
shells $J$ and $H$ depicted in Fig.~\ref{fig:domaindecomp3} overlap
multiple blocks and cylinders, and for this overlap the issue of
gluing is complicated. Since the issue is essentially the same for the
gluing of $H$ to blocks $C,D$ and cylinders 3,4,5 or $J$ to blocks
$B,C$ and cylinders 1,2,3, let us here focus on the first case. The
issue here is that parts of the outer boundary $\partial H^+$ of $H$
sit in blocks $C,D$ and cylinders 3,4,5. Let $\#$ represent one of the
tags $C,D,3,4,5$, and let us consider the portion $\partial H^+_\#$ of
$\partial H^+$ which intersects subdomain $\#$. At nodal points on
$\partial H^+_\#$ we require that the
values of $\psi$ agree whether they are computed with the spectral
representation $\widetilde{\boldsymbol{\psi}}{}^H$ for $H$ or the
spectral representation $\widetilde{\boldsymbol{\psi}}{}^\#$ for
$\#$. For nodal points $(\theta_j,\phi_k)$ this condition is
[cf.~Eq.~\eqref{eq:innershellBCs}]
\begin{equation}
h^+_{jk}\equiv  h^+(\theta_j,\phi_k) = 
\mathcal{P}\psi^\#(\boldsymbol{x}(r_\mathrm{max},\theta_k,\phi_j))
\text{\ \ for } (r_\mathrm{max},\theta_j,\phi_k) \in \partial H^+_\#.
\end{equation}
Here $\mathcal{P}\psi^\#$ is the numerical solution ($\mathcal{P}$
indicates finite expansion) associated with 
$\widetilde{\boldsymbol{\psi}}{}^\#$, and $\boldsymbol{x}$ are
the relevant 3D coordinates on subdomain $\#$.
Looping over all of the subdomains $\# = C,D,3,4,5$ defines the
grid function $h^+_{jk}$ at all nodal points of $\partial H^+_\#$.
The explicit matching conditions 
(equivalent to the + case in \eqref{eq:innershellBCs})
can then be realized by expressing the spherical harmonic transform
$\tilde{h}^+_{\ell q} = \sum_{j=0}^{N_\theta} \sum_{k=0}^{2N_\theta} 
\mathcal{S}_{\ell q, jk} 
h^+_{jk}$
as a matrix-vector product involving all 
$\widetilde{\boldsymbol{\psi}}{}^\#$. 
The resulting equations 
are placed within empty rows of $\mathcal{M}$ which stretch across 
the $H$-$H$ and $H$-$\#$ super-blocks.

Again let $\#$ represent one of the tags $C,D,3,4,5$. 
Then the boundary 
$\partial\#$ of subdomain $\#$ includes a portion $\partial\#_H$ 
overlapping shell $H$ which gives rise to further gluing relations. 
These equations will be inserted into empty rows of $\mathcal{M}$ which
stretch across the $\#$-$\#$ and $\#$-$H$ super-blocks. 
For concreteness, we consider only the $\# = C$ case. Here the $+$ 
case of \eqref{eq:XYfaceBC} is relevant, although the $\tilde{h}^+_{nm}$ 
now arise as the double $XY$-Chebyshev transform of
\begin{align}
h^+_{jk}            & = \sum_{n=0}^{N_r}
                        \sum_{\ell = 0}^{N_\theta}
                        \widetilde{\psi}{}^H_{\ell 0 n}
                        \overline{P}_{\ell 0}(\cos\theta_{jk})
                        T_n(\xi(r_{jk}))
                        \nonumber \\
                    & + \sum_{n=0}^{N_r}
                        \sum_{\ell = 1}^{N_\theta}
                        \sum_{m=1}^{N_\theta}
                        \overline{P}_{\ell m}(\cos\theta_{jk})
                        \big[
                             \widetilde{\psi}{}^H_{\ell, 2m-1, n}
                             \cos (m\phi_{jk})
                           + \widetilde{\psi}{}^H_{\ell, 2m, n}
                             \sin (m\phi_{jk})
                        \big]
                        T_n(\xi(r_{jk})).
\end{align}
The $H$ point $(r_{jk},\theta_{jk},\phi_{jk})$ corresponds to a 
Chebyshev-Gauss-Lobatto collocation point 
$(X(\xi_j),Y(\eta_k),Z_\mathrm{max})$ along 
the top $XY$-face of block $C$. Again, via a matrix representation 
$\widetilde{h}^+_{nm} = 
\sum_{j=0}^{N_X} \sum_{k=0}^{N_Y} 
\mathcal{F}_{nm,jk}h^+_{jk}$ of the transform, we may express this matching 
condition more directly. As mentioned, these equations will be inserted 
into empty rows of $\mathcal{M}$ which stretch across the $C$-$C$ and 
$C$-$H$ super-blocks.  

\subsection{Gluing of a cylinder to a block} 
Here we sketch either the gluing of block $B$ and cylinder 1, block $C$ and 
cylinder 3, or block $D$ and cylinder 5. We focus on the middle case as
a representative example. This process involves both (a) gluing two $YZ$ and 
two $XZ$ faces of block $C$ to cylinder 3, and (b) gluing the inner 
radial boundary of the cylinder to the block. The process for (a) is 
similar to the gluing described in the last paragraph (in which a face
of $C$ is glued to $H$), and we omit a description. To express the 
matching equations which enforce (b), we first define
\begin{equation}
        q^-_{jk} = \sum_{n=0}^{N_X}
                   \sum_{m=0}^{N_Y}
                   \sum_{p=0}^{N_Z}
                   \widetilde{\psi}{}^C_{nmp}
                   T_n(\xi(X_{jk}))
                   T_m(\eta(Y_{jk}))
                   T_p(\chi(Z_{jk})).
\end{equation}
Here we use the following points:
\begin{equation}
\big(X_{jk},Y_{jk},Z_{jk}\big) = 
\big(X(\rho_\mathrm{min},\phi_j,z_k),
 Y(\rho_\mathrm{min},\phi_j,z_k),
 Z(\rho_\mathrm{min},\phi_j,z_k)\big),
\end{equation}
where $(\rho_\mathrm{min},\phi_j,z_k)$ are nodal points along the
inner radial boundary of cylinder 3. Next, we consider the 
Fourier-Chebyshev transform $\tilde{q}^-_{mp} = \sum_{j=0}^{N_\phi} 
\sum_{k=0}^{N_z} \mathcal{C}_{mp,jk} q^-_{jk}$. In terms of the 
transform the matching equations are
\begin{equation}\label{eq:cylinderBCs2}
\sum_{n=0}^{N_\rho} \widetilde{\psi}{}^3_{mnp} \delta_n^-
= \tilde{q}^-_{mp}.
\end{equation}
These equations are inserted into empty rows of $\mathcal{M}$ 
which stretch across the $3$-$3$ and $3$-$C$ superblocks.

\section{Numerical solution of the 3d HRWE}\label{sec:numericaltests}
Both on single subdomains and on the global 2-center multidomain 
$\mathcal{D}$, this section considers numerical solution of the HRWE 
for the field of two point sources in a circular binary orbit. For 
this problem we have an essentially closed-form exact solution, a
superposition of the fields for two point sources, each point source
in a circular orbit and described by the Li\'enard-Wiechert solution 
(\ref{eq:LWSOLN}) found in the appendix. 
A numerical solution is a collection of modal expansion coefficients; 
however, comparisons with the exact solution are always computed in 
physical space on the nodal grid (or grids in the multidomain case) 
dual to the modal expansion.\footnote{As these nodal grids are 
coarse, the $L_2$ and $L_\infty$ norms reported 
in the tables do not settle down quickly.} All numerical solves are
performed iteratively with preconditioned GMRES \cite{GMRES}, and 
this section also describes the relevant preconditioning (both for 
subdomain solves and for the global multidomain solve).

Using the sparse representations described in  
Sec.~\ref{sec:sparse3dHRWE}, in Sec.~\ref{subsec:subdomaintests} we
numerically solve the HRWE on the following subdomains 
(cf.~Fig.~\ref{fig:domaindecomp3} and Table~\ref{table:domaindecomp3}): 
the outer shell $O$, (inner) spherical shell $J$, (inner) cylindrical 
shell $5$, and (inner) 
block $D$. For each subdomain labeled (inner) the HRWE operator 
is implemented as a matrix-vector multiply within preconditioned 
GMRES without restarts. For these subdomain solves, Dirichlet 
boundary conditions are taken from the exact Li\'enard-Wiechert 
solution, but the outer shell also problem involves the
radiation boundary conditions given in Eq.~\eqref{eq:rmaxuBCs}.
The particular subdomains considered in 
Sec.~\ref{subsec:subdomaintests} are representative, 
and similar experimentation on each subdomain has determined the 
chosen truncations for the 2-center multidomain tests described 
in  Sec.~\ref{subsec:multidomain}. Such experiments empirically yield
appropriate truncations necessary to achieve a desired accuracy.
All tests in Secs.~\ref{subsec:subdomaintests} and 
\ref{subsec:multidomain} involve the following configuration: 
two charges, one with $z_H = 1$, $Q_H = 1$ and the other with 
$z_J = -0.9$, $Q_J = 0.5$. Sec.~\ref{subsec:subdomaintests}
considers $\Omega = 0.1, 0.3, 0.5, 0.7$. The rotation rate 
$\Omega = 0.3$ is large for an astrophysical problem,  while 
$\Omega = 0.5, 0.7$ are very large rates chosen to 
``break" our numerical methods.
\begin{figure}[t]
\begin{center}
\subfigure[$\;$Cylinder experiment.]{
\includegraphics[scale=0.325,clip=true,trim=1.0cm 3.5cm 1.0cm 2.5cm]{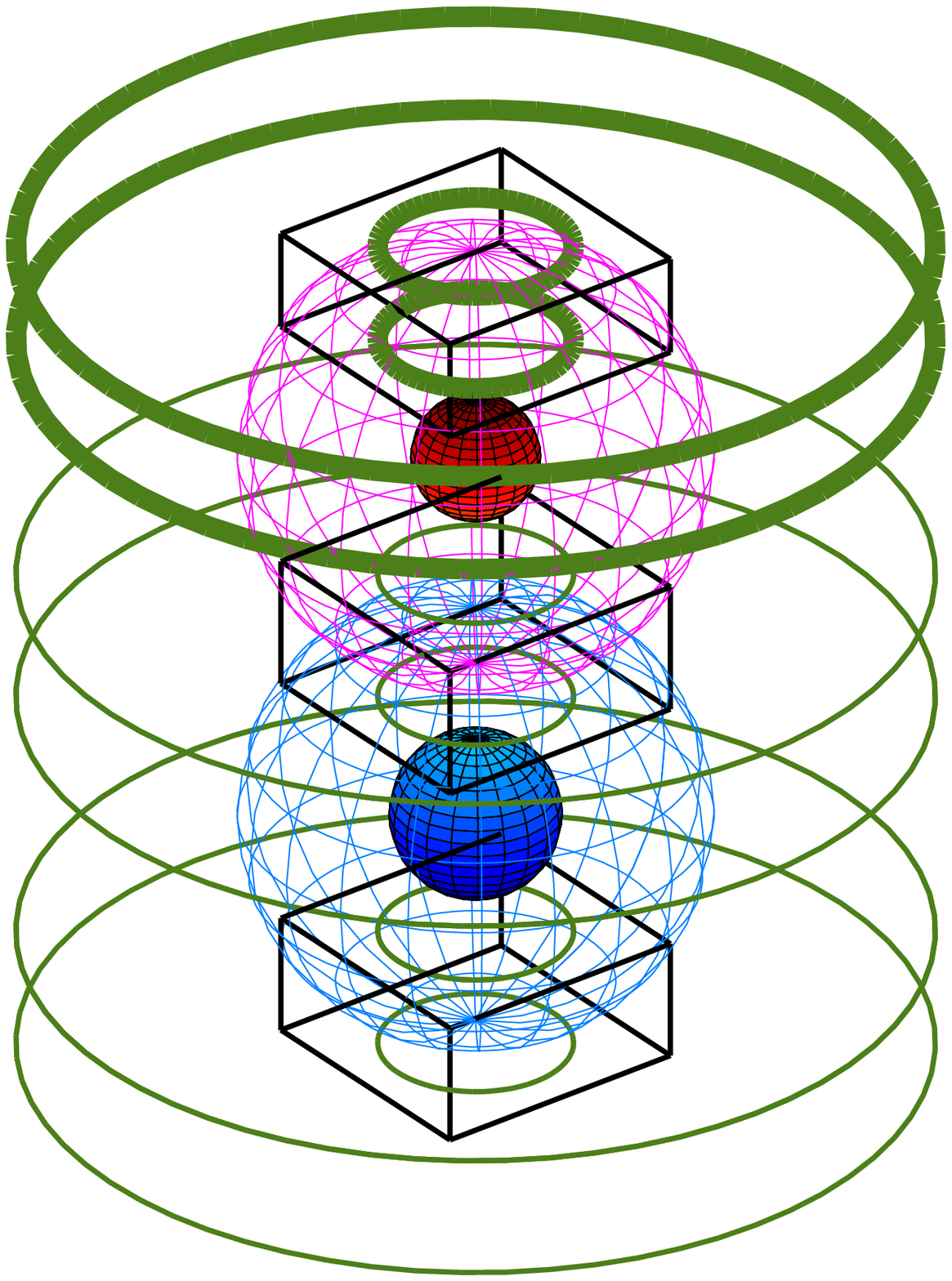}}
\subfigure[$\;$Block experiment.]{
\includegraphics[scale=0.325,clip=true,trim=1.0cm 3.5cm 1.0cm 2.5cm]{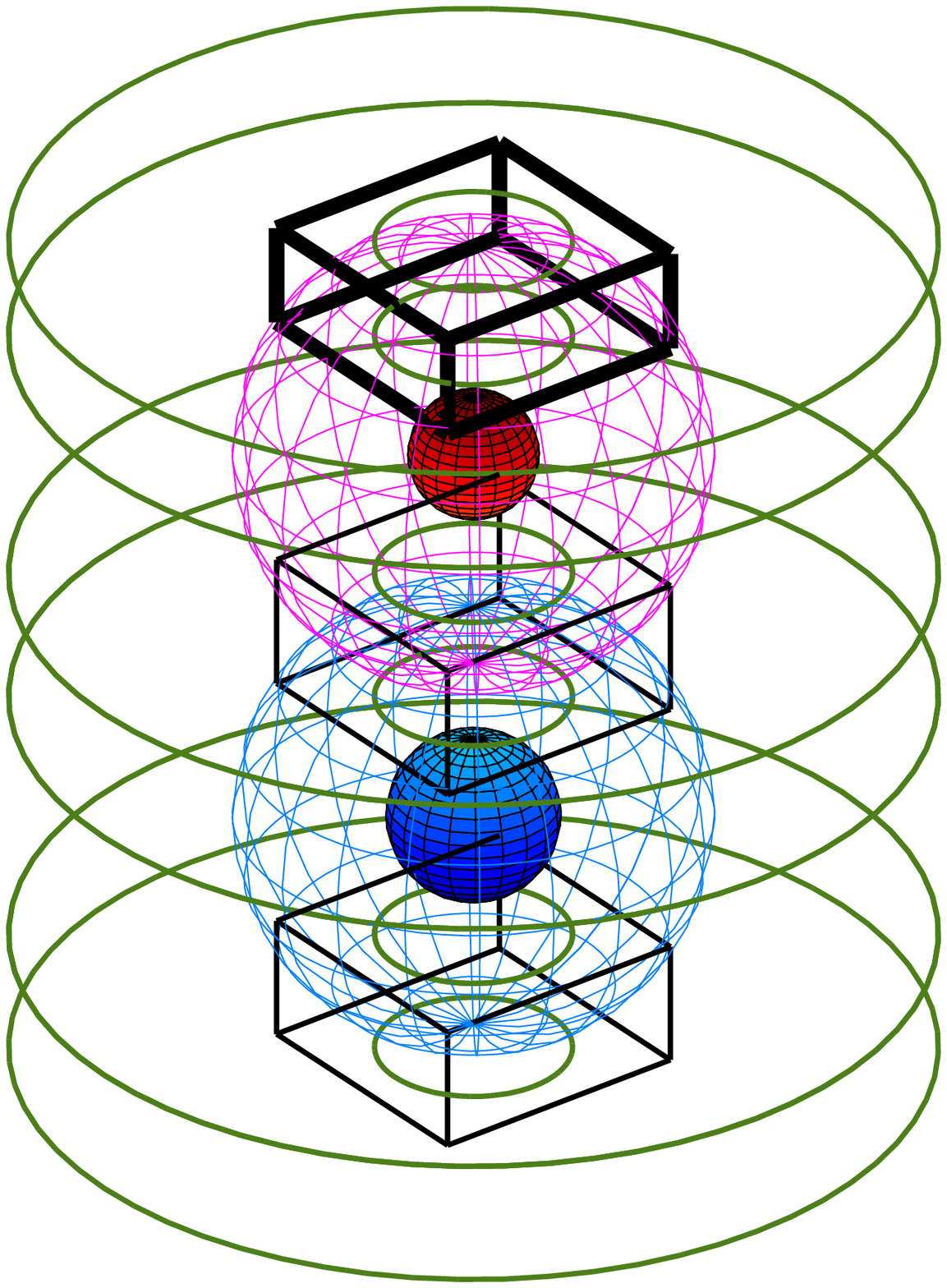}}
\caption{\label{fig:subdomains}
{\sc Individual subdomains.} \em
We consider cylindrical shell $5$ highlighted in the left figure,
block $D$ highlighted in the right, and the bottom
inner spherical shell $J$ shown in both.}
\end{center}
\end{figure}

\subsection{Numerical solution on individual 
subdomains}\label{subsec:subdomaintests}
We consider the outer shell first, since results for this subdomain 
are the most disappointing. The solve for this subdomain differs from 
the rest. Indeed, since the representation of the HRWE on the outer
shell is comprised of $(\ell,m)$ blocks along the block diagonal, we 
invert each of these physical modes using $LU$-factorization
[the ``physical modes" are those {\em not} annihilated 
by the projection operator $\mathbb{P}$ defined in
\eqref{eq:shellvectorsubspace}].
Let $N_\theta = \ell_\mathrm{max}$, so that $\mathcal{N} = 
(N_\theta + 1)(2N_\theta + 1)(N_r+1)$
is the system size, with $\mathcal{N}^2 \sim 4 N_\theta^4 N_r^2$
the storage requirement for the full coefficient matrix. However,
storage of all blocks involves $(N_\theta+1)$ matrices of size 
$(N_r+1)$-by-$(N_r+1)$, one for each zero mode, in addition to 
$\frac{1}{2}N_\theta (N_\theta + 1)$ matrices of size 
$2(N_r+1)$-by-$2(N_r+1)$, one for each fixed-$m$ cos/sin pair. 
Therefore, storage for this solve scales like
\begin{equation}
(N_\theta+1)(N_r+1)^2 + 2N_\theta (N_\theta + 1)(N_r+1)^2
\sim 2 N_\theta^2 N_r^2 = O(N_r \cdot \mathcal{N}).
\end{equation}
Table \ref{table:outershell} collects results for the outer 
shell experiment. While excellent for $\Omega = 0.1$, they
exhibit marked degradation as $\Omega$ increases. A splitting of 
the single outer shell into multiple concentric shells would 
likely yield improved accuracy for large $\Omega$.
\begin{table}\scriptsize
\begin{tabular}{|c|c|c|c|c|c|}
\hline
\multicolumn{6}{|c|}{$\Omega = 0.1$}\\
\hline
\hline
$N_r$ & $\ell_\mathrm{max}$ & $L_2$ error & $L_2$ norm
& $L_\infty$ error & $L_\infty$ norm \\
\hline
65 & 10 & 1.1899E-05 & 1.8977E-01 & 2.7970E-04 & 1.1808E+00\\
85 & 18 & 2.1320E-07 & 1.8725E-01 & 3.6548E-06 & 1.1810E+00\\
125 & 28 & 2.4580E-10 & 1.8472E-01 & 5.4193E-09 & 1.1810E+00\\
185 & 42 & 2.2846E-13 & 1.8297E-01 & 2.7858E-12 & 1.1810E+00\\
\hline
\multicolumn{6}{c}{}\\
\hline
\multicolumn{6}{|c|}{$\Omega = 0.3$}\\
\hline
\hline
$N_r$ & $\ell_\mathrm{max}$ & $L_2$ error & $L_2$ norm
& $L_\infty$ error & $L_\infty$ norm \\
\hline
65 & 10 & 7.4142E-04 & 1.9030E-01 & 1.0557E-02 & 1.2624E+00\\
85 & 18 & 9.4481E-04 & 1.8777E-01 & 1.3180E-02 & 1.2628E+00\\
125 & 28 & 1.2701E-04 & 1.8523E-01 & 2.0861E-03 & 1.2628E+00\\
185 & 42 & 5.1221E-06 & 1.8347E-01 & 1.2307E-04 & 1.2628E+00\\
\hline
\multicolumn{6}{c}{}\\
\hline
\multicolumn{6}{|c|}{$\Omega = 0.5$}\\
\hline
\hline
$N_r$ & $\ell_\mathrm{max}$ & $L_2$ error & $L_2$ norm
& $L_\infty$ error & $L_\infty$ norm \\
\hline
65 & 10 & 1.4622E-02 & 1.9196E-01 & 1.9489E-01 & 1.4875E+00\\
85 & 18 & 5.6234E-02 & 1.9726E-01 & 7.8285E-01 & 1.4904E+00\\
125 & 28 & 6.7047E-03 & 1.8689E-01 & 1.4751E-01 & 1.4996E+00\\
185 & 42 & 3.6457E-03 & 1.8503E-01 & 7.6831E-02 & 1.5055E+00\\
\hline
\multicolumn{6}{c}{}\\
\hline
\multicolumn{6}{|c|}{$\Omega = 0.7$}\\
\hline
\hline
$N_r$ & $\ell_\mathrm{max}$ & $L_2$ error & $L_2$ norm
& $L_\infty$ error & $L_\infty$ norm \\
\hline
65 & 10 & 1.8330E-01 & 2.6488E-01 & 4.2561E+00 & 4.9750E+00\\
85 & 18 & 3.8033E-02 & 1.9502E-01 & 7.1574E-01 & 2.0804E+00\\
125 & 28 & 3.9821E-02 & 1.9408E-01 & 1.0589E+00 & 2.2504E+00\\
185 & 42 & 2.7380E-02 & 1.9029E-01 & 7.3279E-01 & 2.2182E+00\\
\hline
\end{tabular}
\caption{{\sc Outer spherical shell $O$ test}
\label{table:outershell}}
\end{table}

We next consider the inner spherical shell $J$.
Note that $\Omega = 0.5$ corresponds to a shell just within the 
elliptic region, 
but $\Omega = 0.7$ corresponds to a shell which does not lie 
fully within the elliptic region.  Tables \ref{table:shellNoPC} and
\ref{table:shellPC} list errors, without and with preconditioning. 
For the sake of comparison, in both these and subsequent tables
we have chosen the same requested tolerances (for the GMRES solve) 
uniformly in $\Omega$, although for larger $\Omega$ the achieved 
accuracy could likely be attained with a weaker tolerance and fewer 
iterations. The chosen preconditioner is block-Jacobi. Namely, we 
invert physical $(\ell,m)$ modes along the block diagonal using a 
precomputed $LU$-factorization. The storage and scaling properties 
for this preconditioner are exactly the same as described for the 
direct solve on the outer shell.  However, for inner shells the 
HRWE representation is not block diagonal in $(\ell,m)$ pairs 
(as on the outer shell), rather the operator has 
significant bandwidth in both indices. Therefore, storage of 
the full matrix for an inner shell
would require correspondingly larger memory relative to the
preconditioner storage. Further, the preconditioner storage
requirement could be reduced by inverting each sin/cos block mode
independently. Moreover, were the preconditioner chosen to correspond
to only the Laplacian part of the operator, then it could be used for
the solves on both shells if their dimensions and truncations were the
same. In any case, the chosen preconditioner notably
improves the convergence of the GMRES solver.
\begin{table}\scriptsize
\begin{tabular}{|c|c|c|c|c|c|c|c|}
\hline
\multicolumn{8}{|c|}{$\Omega = 0.1$}\\
\hline
\hline
$N_r$ & $\ell_\mathrm{max}$ & $L_2$ error & $L_2$ norm
& $L_\infty$ error & $L_\infty$ norm & iterations & tolerance \\
\hline
12 & 12 & 3.4702E-06 & 1.3516E+00 & 2.2656E-05 & 1.9143E+00 & 54   & 1.0000E-07\\
18 & 23 & 9.9814E-09 & 1.3499E+00 & 3.6488E-08 & 1.9168E+00 & 129  & 1.0000E-09\\
20 & 33 & 6.0107E-11 & 1.3498E+00 & 2.6944E-10 & 1.9173E+00 & 238  & 1.0000E-11\\
30 & 46 & 6.2864E-13 & 1.3481E+00 & 2.8333E-12 & 1.9176E+00 & 415  & 1.0000E-13\\
\hline
\multicolumn{8}{c}{}\\
\hline
\multicolumn{8}{|c|}{$\Omega = 0.3$}\\
\hline
\hline
$N_r$ & $\ell_\mathrm{max}$ & $L_2$ error & $L_2$ norm
& $L_\infty$ error & $L_\infty$ norm & iterations & tolerance \\
\hline
12 & 12 & 7.6593E-06 & 1.3566E+00 & 7.8541E-05 & 1.9255E+00 & 57   & 1.0000E-07\\
18 & 23 & 2.1272E-08 & 1.3550E+00 & 3.8188E-07 & 1.9278E+00 & 138  & 1.0000E-09\\
20 & 33 & 1.1806E-10 & 1.3550E+00 & 2.4707E-09 & 1.9283E+00 & 255  & 1.0000E-11\\
30 & 46 & 4.3184E-13 & 1.3532E+00 & 5.2289E-12 & 1.9286E+00 & 442  & 1.0000E-13\\
\hline
\multicolumn{8}{c}{}\\
\hline
\multicolumn{8}{|c|}{$\Omega = 0.5$}\\
\hline
\hline
$N_r$ & $\ell_\mathrm{max}$ & $L_2$ error & $L_2$ norm
& $L_\infty$ error & $L_\infty$ norm & iterations & tolerance \\
\hline
12 & 12 & 2.3588E-05 & 1.3782E+00 & 2.7337E-04 & 1.9561E+00 & 61   & 1.0000E-07\\
18 & 23 & 8.3249E-08 & 1.3769E+00 & 1.5539E-06 & 1.9579E+00 & 146  & 1.0000E-09\\
20 & 33 & 4.9181E-10 & 1.3770E+00 & 1.1765E-08 & 1.9583E+00 & 271  & 1.0000E-11\\
30 & 46 & 7.9161E-13 & 1.3752E+00 & 1.8530E-11 & 1.9585E+00 & 471  & 1.0000E-13\\
\hline
\multicolumn{8}{c}{}\\
\hline
\multicolumn{8}{|c|}{$\Omega = 0.7$}\\
\hline
\hline
$N_r$ & $\ell_\mathrm{max}$ & $L_2$ error & $L_2$ norm
& $L_\infty$ error & $L_\infty$ norm & iterations & tolerance \\
\hline
12 & 12 & 4.7252E-04 & 1.4267E+00 & 3.1710E-03 & 2.2326E+00 & 87   & 1.0000E-07\\
18 & 23 & 1.1589E-05 & 1.4259E+00 & 1.2079E-04 & 2.2356E+00 & 461  & 1.0000E-09\\
20 & 33 & 8.3186E-08 & 1.4262E+00 & 6.8077E-07 & 2.2353E+00 & 1459 & 1.0000E-11\\
30 & 46 & 1.3484E-09 & 1.4243E+00 & 1.4001E-08 & 2.2351E+00 & 5215 & 1.0000E-13\\
\hline
\end{tabular}
\caption{{\sc Inner spherical shell test $J$ without preconditioning.}
\label{table:shellNoPC}
}
\end{table}
\begin{table}\scriptsize
\begin{tabular}{|c|c|c|c|c|c|c|c|}
\hline
\multicolumn{8}{|c|}{$\Omega = 0.1$}\\
\hline
\hline
$N_r$ & $\ell_\mathrm{max}$ & $L_2$ error & $L_2$ norm
& $L_\infty$ error & $L_\infty$ norm & iterations & tolerance \\
\hline
12 & 12 & 3.4196E-06 & 1.3516E+00 & 2.2499E-05 & 1.9143E+00 & 3    & 1.0000E-07\\
18 & 23 & 3.4877E-09 & 1.3499E+00 & 3.6560E-08 & 1.9168E+00 & 4    & 1.0000E-09\\
20 & 33 & 1.3949E-11 & 1.3498E+00 & 2.1394E-10 & 1.9173E+00 & 4    & 1.0000E-11\\
30 & 46 & 3.0104E-14 & 1.3481E+00 & 3.0975E-13 & 1.9176E+00 & 5    & 1.0000E-13\\
\hline
\multicolumn{8}{c}{}\\
\hline
\multicolumn{8}{|c|}{$\Omega = 0.3$}\\
\hline
\hline
$N_r$ & $\ell_\mathrm{max}$ & $L_2$ error & $L_2$ norm
& $L_\infty$ error & $L_\infty$ norm & iterations & tolerance \\
\hline
12 & 12 & 7.6487E-06 & 1.3566E+00 & 7.8536E-05 & 1.9255E+00 & 4    & 1.0000E-07\\
18 & 23 & 2.0798E-08 & 1.3550E+00 & 3.8215E-07 & 1.9278E+00 & 6    & 1.0000E-09\\
20 & 33 & 1.0362E-10 & 1.3550E+00 & 2.4686E-09 & 1.9283E+00 & 7    & 1.0000E-11\\
30 & 46 & 1.0361E-13 & 1.3532E+00 & 3.1604E-12 & 1.9286E+00 & 9    & 1.0000E-13\\
\hline
\multicolumn{8}{c}{}\\
\hline
\multicolumn{8}{|c|}{$\Omega = 0.5$}\\
\hline
\hline
$N_r$ & $\ell_\mathrm{max}$ & $L_2$ error & $L_2$ norm
& $L_\infty$ error & $L_\infty$ norm & iterations & tolerance \\
\hline
12 & 12 & 2.3577E-05 & 1.3782E+00 & 2.7330E-04 & 1.9561E+00 & 7    & 1.0000E-07\\
18 & 23 & 8.3163E-08 & 1.3769E+00 & 1.5545E-06 & 1.9579E+00 & 10   & 1.0000E-09\\
20 & 33 & 4.9053E-10 & 1.3770E+00 & 1.1755E-08 & 1.9583E+00 & 13   & 1.0000E-11\\
30 & 46 & 6.2389E-13 & 1.3752E+00 & 1.8454E-11 & 1.9585E+00 & 17   & 1.0000E-13\\
\hline
\multicolumn{8}{c}{}\\
\hline
\multicolumn{8}{|c|}{$\Omega = 0.7$}\\
\hline
\hline
$N_r$ & $\ell_\mathrm{max}$ & $L_2$ error & $L_2$ norm
& $L_\infty$ error & $L_\infty$ norm & iterations & tolerance \\
\hline
12 & 12 & 4.7262E-04 & 1.4267E+00 & 3.1713E-03 & 2.2326E+00 & 30   & 1.0000E-07\\
18 & 23 & 1.1596E-05 & 1.4259E+00 & 1.2093E-04 & 2.2356E+00 & 154  & 1.0000E-09\\
20 & 33 & 8.3367E-08 & 1.4262E+00 & 6.8017E-07 & 2.2353E+00 & 429  & 1.0000E-11\\
30 & 46 & 1.3430E-09 & 1.4243E+00 & 1.3999E-08 & 2.2351E+00 & 1390 & 1.0000E-13\\
\hline
\end{tabular}
\caption{{\sc Inner spherical shell $J$ test with preconditioning.}
\label{table:shellPC}
}
\end{table}

Table \ref{table:cylinder} list the results for the corresponding single
cylinder experiment, with block $LU$--preconditioning similar to before.
That is, for each Fourier mode we invert the associated diagonal block.
Our choice \eqref{eq:dpFORcylinders} of direct product structure 
for the cylinders determines that each block is 
$(N_\rho+1)(N_Z+1)$-by-$(N_\rho+1)(N_Z+1)$.
For cylinders, preconditioning amounts to direct inversion of each Fourier 
mode along the block diagonal. With $\mathcal{N} = 
(N_\phi+1)(N_\rho+1)(N_z+1)$ the system size, the storage requirement 
for the preconditioner requires $N_\phi+1$ matrices of size 
$(N_\rho+1)(N_z+1)$-by-$(N_\rho+1)(N_z+1)$, and so scales like	
so
\begin{equation}
(N_\rho+1)^2(N_z+1)^2 (N_\phi+1) = O(N_\rho N_z\cdot \mathcal{N}).
\end{equation}
While $N_\rho N_z \mathcal{N} < \mathcal{N}^2$, this requirement 
is somewhat memory intensive. However, we have observed essentially 
the same performance when using the corresponding Laplacian part of 
the operator to define the preconditioner. Provided that 
the dimensions and truncations of two individual cylinders match, 
the same preconditioner could then be used for both. 

Table \ref{table:block} list 
errors for the block experiment, and again with a block-Jacobi 
preconditioner. In this case there are $N_x + 1$ blocks with size
$(N_y + 1)(N_z+1)$-by-$(N_y + 1)(N_z+1)$. Storage of the
block preconditioner therefore scales as
\begin{equation}
(N_x+1)(N_y + 1)^2(N_z+1)^2 = O(N_yN_z \cdot \mathcal{N}).
\end{equation}
Again, were the preconditioner based on the Laplacian part of the
operator, it might be reused for the solves on different blocks.
\begin{table}\scriptsize
\begin{tabular}{|c|c|c|c|c|c|c|c|c|}
\hline
\multicolumn{9}{|c|}{$\Omega = 0.1$}\\
\hline
\hline
$N_r$ & $N_\phi$ & $N_z$ & $L_2$ error & $L_2$ norm
& $L_\infty$ error & $L_\infty$ norm & iterations & tolerance \\
\hline
13 & 5  & 7  & 7.9884E-08 & 9.0116E-01 & 4.6388E-07 & 1.5004E+00 & 3    & 1.0000E-08\\
19 & 9  & 9  & 5.2802E-10 & 8.9887E-01 & 2.7463E-09 & 1.5006E+00 & 4    & 1.0000E-10\\
23 & 13 & 13 & 5.6239E-13 & 8.9775E-01 & 4.4170E-12 & 1.5006E+00 & 5    & 1.0000E-12\\
29 & 19 & 18 & 8.3992E-15 & 8.9680E-01 & 9.3259E-14 & 1.5007E+00 & 6    & 1.0000E-14\\
\hline
\multicolumn{9}{c}{}\\
\hline
\multicolumn{9}{|c|}{$\Omega = 0.3$}\\
\hline
\hline
$N_r$ & $N_\phi$ & $N_z$ & $L_2$ error & $L_2$ norm
& $L_\infty$ error & $L_\infty$ norm & iterations & tolerance \\
\hline
13 & 5  & 7  & 6.2980E-06 & 9.4046E-01 & 3.8531E-05 & 1.5817E+00 & 6    & 1.0000E-08\\
19 & 9  & 9  & 1.2577E-07 & 9.3796E-01 & 5.9139E-07 & 1.5841E+00 & 10   & 1.0000E-10\\
23 & 13 & 13 & 4.9307E-09 & 9.3677E-01 & 4.0773E-08 & 1.5849E+00 & 14   & 1.0000E-12\\
29 & 19 & 18 & 3.2422E-10 & 9.3574E-01 & 1.3965E-09 & 1.5861E+00 & 18   & 1.0000E-14\\
\hline
\hline
\multicolumn{9}{c}{}\\
\hline
\multicolumn{9}{|c|}{$\Omega = 0.5$}\\
\hline
\hline
$N_r$ & $N_\phi$ & $N_z$ & $L_2$ error & $L_2$ norm
& $L_\infty$ error & $L_\infty$ norm & iterations & tolerance \\
\hline
13 & 5  & 7  & 7.3178E-04 & 1.0239E+00 & 3.9733E-03 & 1.7886E+00 & 59   & 1.0000E-08\\
19 & 9  & 9  & 2.1193E-04 & 1.0218E+00 & 1.2227E-03 & 1.8228E+00 & 161  & 1.0000E-10\\
23 & 13 & 13 & 4.0853E-05 & 1.0203E+00 & 1.8564E-04 & 1.8303E+00 & 531  & 1.0000E-12\\
29 & 19 & 18 & 7.0259E-06 & 1.0190E+00 & 7.7578E-05 & 1.8285E+00 & 1576 & 1.0000E-14\\
\hline
\hline
\multicolumn{9}{c}{}\\
\hline
\multicolumn{9}{|c|}{$\Omega = 0.7$}\\
\hline
\hline
$N_r$ & $N_\phi$ & $N_z$ & $L_2$ error & $L_2$ norm
& $L_\infty$ error & $L_\infty$ norm & iterations & tolerance \\
\hline
13 & 5  & 7  & 3.2460E+00 & 3.4276E+00 & 2.2390E+01 & 2.1481E+01 & 130   & 1.0000E-08\\
19 & 9  & 9  & 2.3293E-02 & 1.1548E+00 & 1.9059E-01 & 2.5635E+00 & 503   & 1.0000E-10\\
23 & 13 & 13 & 1.1546E+00 & 1.6568E+00 & 9.4853E+00 & 1.1688E+01 & 1420  & 1.0000E-12\\
29 & 19 & 18 & 1.1305E-02 & 1.1512E+00 & 1.1635E-01 & 2.5463E+00 & 10816 & 1.0000E-14\\
\hline
\end{tabular}
\caption{{\sc Cylindrical shell 5 test with preconditioning.}
\label{table:cylinder}
}
\end{table}
\begin{table}\scriptsize
\begin{tabular}{|c|c|c|c|c|c|c|c|c|}
\hline
\multicolumn{9}{|c|}{$\Omega = 0.1$}\\
\hline
\hline
$N_x$ & $N_y$ & $N_z$ & $L_2$ error & $L_2$ norm
& $L_\infty$ error & $L_\infty$ norm & iterations & tolerance \\
\hline
14 & 14 & 7  & 3.7513E-07 & 1.1367E+00 & 4.2360E-06 & 1.7854E+00 & 41  & 1.0000E-08 \\
19 & 19 & 9  & 6.3235E-09 & 1.1394E+00 & 1.3616E-07 & 1.8098E+00 & 62  & 1.0000E-10 \\
28 & 28 & 13 & 1.4351E-11 & 1.1418E+00 & 3.0822E-10 & 1.8040E+00 & 102 & 1.0000E-12 \\
32 & 32 & 18 & 1.2749E-13 & 1.1421E+00 & 5.3182E-12 & 1.8054E+00 & 141 & 1.0000E-14 \\
\hline
\multicolumn{9}{c}{}\\
\hline
\multicolumn{9}{|c|}{$\Omega = 0.3$}\\
\hline
\hline
$N_x$ & $N_y$ & $N_z$ & $L_2$ error & $L_2$ norm
& $L_\infty$ error & $L_\infty$ norm & iterations & tolerance \\
\hline
14 & 14 & 7  & 3.9020E-07 & 1.1955E+00 & 4.2789E-06 & 1.8911E+00 & 42  & 1.0000E-08 \\
19 & 19 & 9  & 6.4495E-09 & 1.1986E+00 & 1.4194E-07 & 1.9176E+00 & 65  & 1.0000E-10 \\
28 & 28 & 13 & 1.4448E-11 & 1.2013E+00 & 4.9319E-10 & 1.9116E+00 & 109 & 1.0000E-12 \\
32 & 32 & 18 & 8.4807E-14 & 1.2017E+00 & 2.9017E-12 & 1.9131E+00 & 154 & 1.0000E-14 \\
\hline
\multicolumn{9}{c}{}\\
\hline
\multicolumn{9}{|c|}{$\Omega = 0.5$}\\
\hline
\hline
$N_x$ & $N_y$ & $N_z$ & $L_2$ error & $L_2$ norm
& $L_\infty$ error & $L_\infty$ norm & iterations & tolerance \\
\hline
14 & 14 & 7  & 7.5596E-07 & 1.3306E+00 & 4.0594E-06 & 2.1990E+00 & 77   & 1.0000E-08 \\
19 & 19 & 9  & 9.7796E-09 & 1.3351E+00 & 1.6231E-07 & 2.2244E+00 & 266  & 1.0000E-10 \\
28 & 28 & 13 & 2.2103E-10 & 1.3391E+00 & 2.8426E-09 & 2.2217E+00 & 1495 & 1.0000E-12 \\
32 & 32 & 18 & 4.3594E-12 & 1.3399E+00 & 5.1787E-11 & 2.2231E+00 & 3089 & 1.0000E-14 \\
\hline
\multicolumn{9}{c}{}\\
\hline
\multicolumn{9}{|c|}{$\Omega = 0.7$}\\
\hline
\hline
$N_x$ & $N_y$ & $N_z$ & $L_2$ error & $L_2$ norm
& $L_\infty$ error & $L_\infty$ norm & iterations & tolerance \\
\hline
14 & 14 & 7  & 4.0731E-03 & 1.5016E+00 & 1.9997E-02 & 3.0390E+00 & 454  & 1.0000E-08 \\
19 & 19 & 9  & 3.9332E-04 & 1.5065E+00 & 2.1243E-03 & 3.1084E+00 & 1349 & 1.0000E-10 \\
28 & 28 & 13 & 2.3057E-06 & 1.5118E+00 & 1.3776E-05 & 3.1019E+00 & 5337 & 1.0000E-12 \\
32 & 32 & 18 & 1.0375E-06 & 1.5133E+00 & 6.6072E-06 & 3.0967E+00 & 20000${}^*$ & 1.0000E-14\\
\hline
\end{tabular}
\caption{{\sc Block $D$ test with preconditioning.} The asterisk on 
$20000^*$ indicates the convergence was halted before the tolerance 
had been achieved; the achieved tolerance $1.3\times 10^{-14}$ was
close to that requested.
\label{table:block}
}
\end{table}

\subsection{Numerical solution on the 2-center multidomain}\label{subsec:multidomain}
We have also used GMRES \cite{GMRES} to solve 
the linear system $\mathcal{M}\widetilde{\mathbf{\Psi}} = 
 \mathcal{B}\widetilde{\mathcal{G}}$
given in Eq.~(\ref{eq:2centerLinearSystem}) and corresponding
to the HRWE on the full 2-center multidomain $\mathcal{D}$.
Section \ref{sec:sparse3dHRWE} has described the coefficient 
matrix $\mathcal{M}$, and therefore also implementation of 
the ``matrix-vector multiply" $\widetilde{\mathbf{\Psi}} 
\rightarrow \mathcal{M}\widetilde{\mathbf{\Psi}}$. 
Implementation of this multiply is required by the GMRES 
algorithm (with or without preconditioning). However, 
a simple {\em unpreconditioned} GMRES strategy results in
extremely poor convergence. Therefore, we have implemented
(left) {\em preconditioned} GMRES which further requires
implementation of the operation $\widetilde{\mathbf{\Psi}}
\rightarrow \mathcal{M}_\mathrm{approx}^{-1}
\widetilde{\mathbf{\Psi}}$ in terms of a suitable 
approximate inverse $\mathcal{M}_\mathrm{approx}^{-1} 
\simeq \mathcal{M}^{-1}$. In this section we describe 
application of $\mathcal{M}_\mathrm{approx}^{-1}$, 
and document tests of the full global solve. We stress that 
the preconditioning afforded by 
$\mathcal{M}_\mathrm{approx}^{-1}$ is neither (i) 
the integration ``preconditioning" technique used to achieve 
sparse representations of (\ref{eqn:HRWEoperator}) on each 
of the basic subdomains nor (ii) the preconditioning
(typically a form of block-$LU$) used for individual subdomain
solves. However, type (ii) preconditioning does define part of 
the $\mathcal{M}_\mathrm{approx}^{-1}$ application.

\begin{figure}[t]
\begin{center}
\subfigure[$\;$Inner shells $J$ and $H$.]{
\includegraphics[scale=0.45,clip=true,trim=9cm    1.25cm 9cm 0]{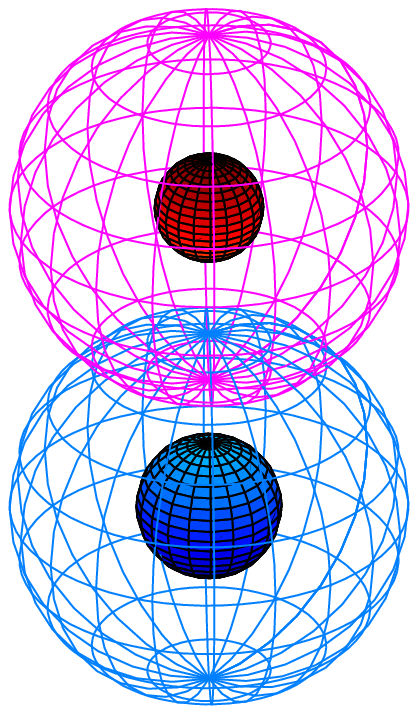}}
\subfigure[$\;$Glued subregion $R$.]{
\includegraphics[scale=0.45,clip=true,trim=8cm    1cm 8cm 0]{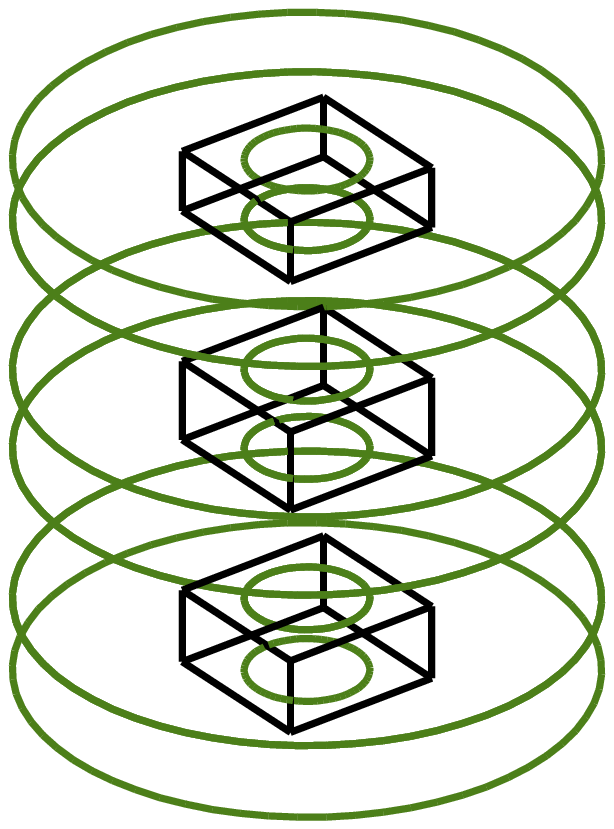}}
\subfigure[$\;$Outer shell $O$.]{
\includegraphics[scale=0.725,clip=true,trim=0.5cm 2.25cm 1cm 0]{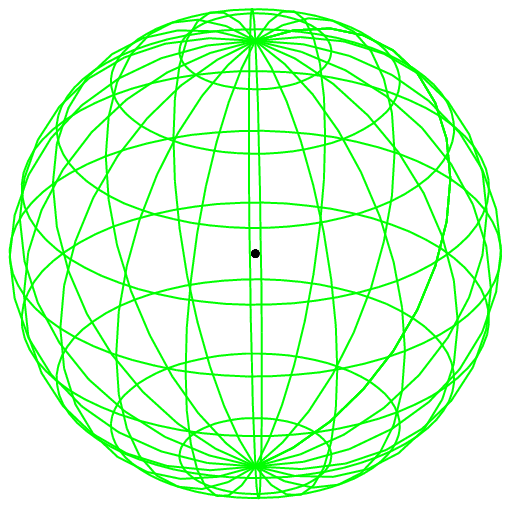}}
\caption{\label{fig:blkscyls}
{\sc Alternating Schwarz preconditioner.} Numerical
solution of the HRWE each subdomain/subregion above
defines the preconditioner. Boundary conditions for 
the solves are obtained through subdomain/subregion 
interpolation as described in the text. For the outer
shell shown in (c), the small dot in the center is,
to scale, the inner configuration comprised of (a) 
and (b).}
\end{center}
\end{figure}

The action of $\mathcal{M}_\mathrm{approx}^{-1}$ is defined through
the simple alternating Schwarz method \cite{SBG1996}. Application 
of this preconditioner relies on independent numerical solves over 
(i) the inner shells $J$ and $H$, (ii) the glued 
subregion\footnote{Whereas the basic spectral elements (such as 
shell $J$, block $B$, and cylinder 1) have been called {\em subdomains},
we informally refer to the multidomains $R$ and $G$ 
(defined later) as a {\em subregions}.} $R$
comprised of blocks and cylinders depicted in 
Fig.~\ref{fig:blkscyls}, and (iii) the outer spherical shell 
$O$. More precisely, starting with a vanishing initial 
vector $\widetilde{\mathbf{\Psi}}$ we perform the following 
iteration.
\begin{itemize}
\item[1.] Solve (also by GMRES, as described in 
Sec.~\ref{subsec:subdomaintests}) the HRWE on the inner shells 
$J$ and $H$. For these solves inner Dirichlet boundary 
conditions are the fixed physical ones, while outer boundary 
conditions stem from interpolation of the numerical solution 
on $R$ (which is initially zero). The tolerance for these 
solves is typically {\tt 0.1*tol}, where {\tt tol} is 
the tolerance for the global GMRES solve of 
$\mathcal{M}\widetilde{\mathbf{\Psi}} =
\mathcal{B}\widetilde{\mathcal{G}}$.

\item[2.] Solve (also by GMRES) the HRWE on $R$. For this 
solve inner Dirichlet boundary conditions stem from interpolation 
of the solutions on $J$ and $H$, while outer Dirichlet boundary 
conditions stem from interpolation of the solution on the outer 
shell $O$ (which is initially zero). This GMRES solve must also 
be preconditioned, as discussed shortly. The tolerance for 
this solve is typically {\tt 0.2*tol}.

\item[3.] Solve the HRWE on the outer spherical shell $O$, 
with inner Dirichlet boundary conditions stemming from 
interpolation of the numerical solution on $R$ and the outer 
radiation boundary conditions described in 
Sec.~\ref{subsec:outershell}. As described in 
Sec.~\ref{subsec:subdomaintests}, this solve is 
performed via direct block-by-block $LU$ factorization 
(note that the factorization of each block mode is 
{\em precomputed} and then used over and over in this 
third step).
\end{itemize}
This three-step iteration may be viewed as the 
Gauss-Seidel method, here applied in block form to 
$J\bigcup H$, $R$, $O$. Typically, we have chosen 4 
sweeps of this block Gauss-Seidel method. 
Step 2 requires its own preconditioning to enhance
convergence. Here we have again employed the alternating 
Schwarz method, now with blocks corresponding to $B$, 
$C$, $D$, and the subregion $G$ which is the composite 
of glued cylinders (1-5). This ``inner" preconditioning 
typically involves 5 sweeps, with appropriate interpolation. 
Each individual GMRES solve on $B$, $C$, $D$, and $G$ uses 
the tolerance {\tt 0.1*tol}. Table~\ref{table:PC} depicts
the overall multilevel preconditioning scheme.
\begin{table}
\begin{center}
\framebox{
\setlength{\unitlength}{.01in}
\begin{picture}(575,220)
\thicklines
\put(0,200){\line(1,0){20}}
\put(30,200){\makebox(20,0){$\mathcal{D}$}}
\put(65,200){\makebox(275,0)[l]{(GMRES solve, alternating Schwarz method as PC)}}
\put(0,115){\line(0,1){85}}
\put(0,165){\vector(1,0){20}}
\put(30,165){\makebox(20,0){$J,H$}}
\put(90,165){\makebox(275,0)[l]{(GMRES solve with block-$LU$ PC)}}
\put(0,140){\vector(1,0){20}}
\put(75,140){\line(-1,0){20}}
\put(30,140){\makebox(20,0){$R$}}
\put(90,140){\makebox(275,0)[l]{(GMRES solve, alternating Schwarz method as PC)}}
\put(0,115){\vector(1,0){20}}
\put(30,115){\makebox(20,0){$O$}}
\put(90,115){\makebox(275,0)[l]{(direct block-$LU$ solve)}}
\put(445,170){\line(1,0){20}}
\put(445,110){\line(1,0){20}}
\put(465,110){\line(0,1){60}}
\put(475,147){\makebox(100,0)[l]{interpolation}}
\put(475,132){\makebox(100,0)[l]{between solves}}
\put(75,15){\line(0,1){125}}
\put(75,90){\vector(1,0){20}}
\put(100,90){\makebox(275,0){$B$ (GMRES solve with block-$LU$ PC)}}
\put(75,65){\vector(1,0){20}}
\put(100,65){\makebox(275,0){$C$ (GMRES solve with block-$LU$ PC)}}
\put(75,40){\vector(1,0){20}}
\put(100,40){\makebox(275,0){$D$ (GMRES solve with block-$LU$ PC)}}
\put(75,15){\vector(1,0){20}}
\put(100,15){\makebox(275,0){$G$ (GMRES solve with block-$LU$ PC)}}
\put(445,95){\line(1,0){20}}
\put(445,10){\line(1,0){20}}
\put(465,10){\line(0,1){85}}
\put(475,60){\makebox(100,0)[l]{interpolation}}
\put(475,45){\makebox(100,0)[l]{between solves}}
\end{picture}}
\end{center}
\caption{\label{table:PC}
{\sc Multilevel preconditioning scheme.}}
\end{table}

Before turning to tests of the full solve, we consider the solve on the
multidomain subregion $G$ comprised of the glued cylinders (1-5). Again, 
this solve is performed as part of the preconditioner for step 2 of the 
global preconditioner (see Table~\ref{table:PC}). Table 
\ref{table:subregionG} collects errors and iteration counts associated 
with this solve for increasing truncations. Each solve documented in
the table has been started with the zero vector as initial iterate,
and here we employ restarting after 20 iterations. The reported
iteration counts in Table \ref{table:subregionG} are cumulative over 
restarts. The individual block-$LU$ preconditioning on each subdomain 
(1-5) is the only preconditioning used for this solve. Nevertheless, it 
suffices to drastically reduce the number of iterations (which would 
otherwise be in the thousands, with or without restarts).
\begin{table}\scriptsize
\begin{tabular}{|c|c|c|c|c|c|c|c|c|c|c|c|c|}
\hline
\multicolumn{13}{|c|}{$\Omega = 0.1$}\\
\hline
\hline
$N_r^1$ & $N_\phi^1$ & $N_z^1$ & $N_z^2$ & $N_z^3$ & $N_z^4$ & $N_z^5$
& $L_2$ error & $L_2$ norm
& $L_\infty$ error & $L_\infty$ norm & iterations & tolerance \\
\hline
13 & 5  & 7  & 17 & 7  & 17 & 7  & 2.2251E-06 & 9.8806E-01 & 2.7804E-05 & 2.4773E+00 & 17 & 1.0000E-06 \\
19 & 9  & 9  & 23 & 9  & 23 & 9  & 3.5812E-08 & 1.0077E+00 & 2.2023E-07 & 2.4781E+00 & 22 & 1.0000E-08 \\
23 & 13 & 14 & 31 & 16 & 31 & 14 & 1.2344E-10 & 1.0063E+00 & 1.2046E-09 & 2.4782E+00 & 28 & 1.0000E-10 \\
29 & 19 & 15 & 35 & 15 & 35 & 15 & 6.6478E-12 & 1.0083E+00 & 7.4462E-11 & 2.4783E+00 & 42 & 1.0000E-12 \\
29 & 19 & 18 & 39 & 21 & 39 & 18 & 4.9252E-13 & 1.0063E+00 & 5.3570E-12 & 2.4783E+00 & 45 & 1.0000E-13 \\
\hline
\end{tabular}
\caption{{\sc Solution of the HRWE on the glued cylinder subregion $G$.} 
The reported truncations $N_r^1$ and $N_\phi^1$ were also used 
for cylinders 2,3,4, and 5.
\label{table:subregionG}
}
\end{table}

Results for the full solve appear in Table \ref{table:twocenterD}. 
Notice that the largest truncation involves more than half a 
million unknowns ($597788$ to be precise). In fact the number of 
unknowns is larger, since we add modes to shells, but here count 
only the ``physical modes" for allowable $(\ell,m)$ pairs 
(cf.~Sec.~\ref{subsec:outershell}).
Each solve in the table is used as the initial guess for the next, 
which is why the count of outer GMRES iterations goes down.
\begin{table}\scriptsize
\begin{tabular}{|c|c|c|c|c|c|c|}
\hline
\multicolumn{7}{|c|}{$\Omega = 0.1$}\\
\hline
\hline
MPSPD   & $L_2$ error & $L_2$ norm  & $L_\infty$ error & $L_\infty$ norm & iterations & tolerance \\
\hline
15.7 &  3.7532E-06 &   7.0509E-01 &   9.9579E-05 &   3.6556E+00 &        5 &   1.0000E-05\\
23.9 &  4.2440E-08 &   7.8382E-01 &   5.8222E-07 &   3.6563E+00 &        3 &   1.0000E-07\\
31.0 &  2.6333E-10 &   8.3492E-01 &   4.0406E-09 &   3.6564E+00 &        3 &   1.0000E-09\\
37.2 &  4.1855E-12 &   9.3982E-01 &   8.6696E-11 &   3.6565E+00 &        3 &   1.0000E-11\\
37.9 &  4.7733E-13 &   9.5252E-01 &   8.2254E-12 &   3.6565E+00 &        2 &   1.0000E-12\\
\hline
\end{tabular}
\caption{{\sc Solution of the HRWE on the 2-center multidomain $\mathcal{D}$.} 
Here MPSPD stands for {\em modes per subdomain per dimension}. Note that
an MPSPD of 37.9 corresponds to  $\text{(11 subdomains)} \times (37.9^3) \simeq 599000$
unknowns.
\label{table:twocenterD}
}
\end{table}

\section{Conclusion}\label{sec:conclusion}
We close by summarizing the results of this paper 
and describing our outlook on future work. In both the
summary and description, we discuss both our numerical 
methods and the physical problem we aim to solve. 

\subsection{Results}
We have demonstrated the feasibility of solving a partial 
differential equation in three independent variables by modal 
spectral methods based on the technique of integration 
preconditioning. As designed, the technique yields an 
algorithmic way to achieve a sparse spectral 
formulation of the PDE problem with consistent incorporation 
of boundary conditions. However, particularly in higher 
dimensional settings, an integration ``preconditioner" may
not be an optimal approximate inverse in any known sense; as 
a result the technique would not seem practical in and of itself. 
Here we mean that, for a higher dimensional problem like ours, 
the sole use of integration preconditioning will likely 
lead to prohibitively large iteration counts when using Krylov 
methods and/or loss of accuracy due to poor conditioning. At 
least for our problem, we have demonstrated that both issues 
may be surmounted by {\em further preconditioning}. In 
particular, studying our problem on a given subdomain 
(spectral element), we have empirically demonstrated that 
block Jacobi preconditioning (with each block inverted by 
$LU$ factorization) is effective for the banded matrix 
produced by integration preconditioning. Moreover, for the 
matching of subdomains 
in our multidomain approach the alternating Schwarz method
(an elementary domain decomposition preconditioner) works
well. Given that little seems known about preconditioning 
for modal methods, whereas preconditioning for nodal methods 
is well developed, we believe that our demonstration of 
effective modal preconditioning based on rather standard 
methods is remarkable.

In addition to modal preconditioning, other aspects of our 
work are new from the standpoint of modal spectral methods, 
in particular its multidomain character and focus on a 
mixed-type problem. Ref.~\cite{CHHT} already presented the 
outline for applying the integration preconditioning technique 
to higher dimensional problems, that is to PDEs. While we have 
carried out and presented the details of such an application, 
our work has gone further in developing a 3D
{\em multidomain} version of the technique (Ref.~\cite{LauPriceI}
consider the {\em multidomain} case in 2D). In 
particular, we have presented 
the details of gluing constituent subdomains, and how this 
gluing is reflected in the overall linear system. As another 
new, and unusual, aspect, our work is the first successful 
application of integration preconditioning to a three 
dimensional mixed-type problem, a problem with both elliptic 
and hyperbolic regions. Whence it has numerically confirmed 
once more (cf.~\cite{LauPriceI,BOP2005,BBP2006,HernandezPrice2009}) 
that such problems can be well-posed; see \cite{BicakSchmidt}
for a theoretical discussion.
The use of a multidomain decomposition is of special interest 
for mixed problems like ours, since the type change need not 
occur in all subdomains. Indeed, for our example, it occurs 
on a cylinder that intersects only the outer spherical shell. 
When the nonlinearities of the actual physical problem are 
included, this feature of our domain decomposition may prove 
useful, because the true physical equations will be only 
mildly nonlinear on the outer shell, with the strongest 
nonlinearities confined to subdomains on which the 
equations are elliptic. Our work therefore
suggests that we might treat the outer shell differently from 
the inner subdomains when solving the full nonlinear problem.

\subsection{Outlook}
While we have demonstrated that our mix of methods delivers
efficiency and remarkable accuracy when applied to a nontrivial 
3D model problem, a number of issues merit 
further investigation. These include both particular challenges 
in the application of this paper's methods to helically symmetric 
general relativistic binary fields (the problem of our interest),
and numerical analysis questions pertaining to integration 
preconditioning as a method for more general problems.

The numerical analysis issues center on the value of 
integration preconditioning, or sparsification, in the 
solution of higher dimensional PDEs, particularly in 
the context of a multidomain approach. Here we have applied 
the method to only one linear PDE, with an empirical 
demonstration of its success. For any given linear equation,
a fuller investigation of integration sparsification for 
multidomain scenarios would focus on the interplay between 
condition number, field of values (Rayleigh quotients), and 
computational efficiency (iteration counts). All of these 
issues would be examined both before and after some form 
of ``ordinary'' preconditioning, e.g.~the combination of
block-$LU$ and alternating Schwarz preconditioning used in 
this paper. The sparse matrices produced by integration 
sparsification allow for quicker matrix multiplies in a 
Krylov method like GMRES. Our work suggests that this 
advantage might be gained without large iteration counts,
but the issue deserves more careful study. The efficient 
treatment of nonlinearities is also worthy of 
investigation, and any such study would build upon the 
results already given in Ref.~\cite{CHHT}. At present, we 
are in process of evaluating integration sparsification in 
the context of these issues, mostly with 2D model problems. 

Several challenges remain if we are to apply some variant 
of our method to the problem of helically symmetric general 
relativistic binary fields. First, we must test the 
efficiency of our method in solving a nonlinear HRWE. In 
practice, this should not be a problem.  The strongest 
nonlinearities will occur closest to the black hole 
sources, i.e.~near the surfaces on which the inner boundary 
conditions are set. By choosing these boundaries some 
distance from the sources, we can, at the cost of accuracy 
in mathematically representing the physical problem, 
reduce the severity of the nonlinearities. The real 
question, then, is not whether we can handle nonlinearities, 
but how close to the sources  
the inner boundaries can be placed. 
Second, we must replace the outgoing 
radiative boundary conditions with ``standing wave boundary 
conditions," as described in Ref.~\cite{Andradeetal2004}. 
This change is straightforward in a linearized problem, and, 
as explained in Ref.~\cite{Andradeetal2004}, should not 
pose great difficulty in nonlinear general relativity.
Third, we must move from the scalar problem considered 
here to the actual tensor problem. 
Solution of the helically symmetric problem of a binary in 
full general relativity will require all the information 
in the tensorial fields, and the coupling of those fields.  
This proved to be the greatest challenge for the solution 
method presented in Ref.~\cite{HernandezPrice2009}, and
it severely limited the achievable accuracy. We are
confident that the method described in this paper will
deliver the accuracy needed to find useful solutions.

The methods developed here have been motivated by the problem of 
binary inspiral in general relativity. However, our methods may
find broader use; they might be applied to problems distinct 
from the helically symmetric mixed PDEs of the periodic standing 
wave approximation. As a salient example, multidomain spectral 
methods are already being used in the elliptical problem of 
generating binary black hole initial data 
\cite{Pfeifferthesis,PKSTelliptic}. Our set of methods, with 
integration sparsification, might be used as an alternative
approach.

\section{Acknowledgments}\label{sec:acknowledgments}
We gratefully acknowledge support from
NSF grants PHY 0855678 to the University of New Mexico
and 
PHY 0554367
to the University of Texas at Brownsville. 
For helpful comments and discussions we thank
T.~Hagstrom, J.~Hesthaven, H.~Pfeiffer, G.~von Winckel, 
and particularly E.~Coutsias.

\appendix
\section{Explicit solution for a point source.}
This appendix presents two representations for an exact 
solution to the HRWE, namely the solution for a point 
source in a circular orbit. Superposition of
two such solutions yields the binary field exploited in
 our numerical tests. As before, let 
$(\tilde{x},\tilde{y},{z}=\tilde{z}) = (r\sin\theta\cos\varphi,
r\sin\theta\sin\varphi,r\cos\theta)$ represent the comoving Cartesian
coordinates, where $\varphi = \phi - \Omega t$. In terms of the
comoving coordinates, we define the Laplacian $\tilde{\nabla}{}^2
\equiv \partial^2_{\tilde{x}} + \partial^2_{\tilde{y}} + \partial^2_z$
and $\partial_\varphi$ operators and consider the inhomogeneous HRWE
\begin{equation}\label{eq:AppendixForcedHRWE}
\big(\tilde{\nabla}{}^2 - \Omega^2\partial^2_\varphi)\psi = 
     -4\pi
      \frac{\delta(r-a)}{a^2}
      \delta(\cos\theta)
      \delta(\varphi-\varphi_0),
\end{equation}
where $(a,\pi/2,\varphi_0)$ specifies the location of the source point
in the spherical polar system associated with $(\tilde{x},\tilde{y},z)$.
We set $\varphi_0 = 0$, since this shift can always be reinserted via
the replacement $\varphi \rightarrow \varphi - \varphi_0$ in the 
representations \eqref{eq:seriesSOLN} and \eqref{eq:LWSOLN} given below.
Using standard methods of separation of variables and 
one-dimensional Green's functions, we find the series representation 
for a particular solution to \eqref{eq:AppendixForcedHRWE},
\begin{align}
\psi(\tilde{x},\tilde{y},z)
   & = 2\sum_{\ell = 0}^\infty
       \frac{1}{2\ell+1}
       \overline{P}_{\ell 0}(\cos\theta)
       \overline{P}_{\ell 0}(0)
       \frac{r_{<}^{\ell}}{r_{>}^{\ell+1}}
\nonumber \\
   - 4\Omega &\sum_{\ell = 1}^\infty \sum_{m = 1}^\ell
       m  \overline{P}_{\ell m}(\cos\theta)
       \overline{P}_{\ell m}(0)
       j_{\ell}(m\Omega r_{<})
       \big[n_{\ell}(m\Omega r_{>})\cos(m\varphi)+
            j_{\ell}(m\Omega r_{>})\sin(m\varphi)
       \big].
\label{eq:seriesSOLN}
\end{align}
Here $\overline{P}_{\ell m}(u)$ is a normalized associated
Legendre function, and $j_\ell(z)$ and $n_\ell(z)$ are respectively 
spherical Bessel functions of the first and second kind \cite{AS}.
Moreover, we adopt the standard notations $r_< = \mathrm{min}(a,r)$
and $r_> = \mathrm{max}(a,r)$. The series is poorly convergent near 
$r = a$; however, for $\Omega \ll 1$ it converges rapidly for
$r \gg a$.

To derive a separate representation of the same series which can be used
near $r = a$, we consider the equivalent problem for the 3+1 wave equation
written in inertial, rather than comoving coordinates,
\begin{equation}
\big(\nabla^2 - \partial^2_t)\Psi =
     -4\pi
      \frac{\delta(r-a)}{a^2}
      \delta(\cos\theta)
      \delta(\phi - \Omega t).
\end{equation}
In the inertial frame the source point has the 
time-dependent location
\begin{equation}
\boldsymbol{\xi}(t) = a \cos(\Omega t) \mathbf{e}_x 
                    + a\sin(\Omega t) \mathbf{e}_y.
\end{equation}
Therefore, we wish to find the retarded solution to
\begin{equation}
\big(\nabla^2 - \partial^2_t)\Psi =
     -4\pi \delta^{(3)}(\mathbf{x}-\boldsymbol{\xi}(t)),
\label{MWproblem}
\end{equation}
and then evaluate it at the field point 
\begin{equation}
\mathbf{x}(t) = z \mathbf{e}_z 
           + \rho \cos(\phi + \Omega t) \mathbf{e}_x
           + \rho \sin(\phi + \Omega t) \mathbf{e}_y,
\end{equation}
where $\rho^2 = x^2 + y^2 = \tilde{x}{}^2 + \tilde{y}{}^2$.
Notice that the evaluation point $\mathbf{x}(t)$ rotates with 
the source; whence this latter evaluation will effectively 
remove the time dependence. The retarded-time Green's function 
for the wave operator is 
\begin{equation}
G_\mathrm{ret}(t,\mathbf{x};t',\mathbf{x}') = \frac{1}{4\pi}
\frac{\delta(t - t' 
- |\mathbf{x}-\mathbf{x}'|)}{|\mathbf{x}-\mathbf{x}'|},
\end{equation}
and it obeys
\begin{equation}
\big(\nabla^2 - \partial^2_t)
G_\mathrm{ret}(t,\mathbf{x};t',\mathbf{x}')
     = -\delta(t-t')
        \delta^{(3)}(\mathbf{x}-\mathbf{x'}).
\end{equation}
We obtain the desired solution to (\ref{MWproblem}) via spacetime 
convolution of $4\pi \delta^{(3)}(\boldsymbol{\xi}(t))$ 
with $G_\mathrm{ret}(t,\mathbf{x};t',\mathbf{x}')$. The details of 
this calculation are given in the textbook by Matthews and Walker 
\cite{MW}, and the result is
\begin{equation}
\Psi(t,\mathbf{x}) = 
\frac{1}{|\mathbf{x}-\boldsymbol{\xi}(t')|
       -\dot{\boldsymbol{\xi}}(t')\cdot
        (\mathbf{x}-\boldsymbol{\xi}(t'))}\,,
\label{eq:MWresult}
\end{equation}
where $\dot{\boldsymbol{\xi}}$ denotes the derivative of 
${\boldsymbol{\xi}}(t')$ with respect to its argument.
Here $t'$ is retarded time which obeys 
$t - t' = |\mathbf{x}-\boldsymbol{\xi}(t')|$. Setting
$\mathbf{x} = \mathbf{x}(t)$, we find
\begin{align}
(t-t') & = \big[z^2+\rho^2+a^2
       -2 a\rho\cos(\phi -\Omega t')\big]^{1/2} \nonumber\\
       & = \big[z^2+\rho^2+a^2
       -2 a\rho\cos(\varphi +\Omega (t- t'))\big]^{1/2}.
\label{tmt'}
\end{align}
With $(\rho,z,\varphi)$ near $(a,0,0)$, we may use the last 
formula to  numerically compute $t-t'$ via fixed-point 
iteration. Finally, since $\psi(\tilde{x},\tilde{y},z) = 
\Psi(t,\mathbf{x}(t))$, we then have
\begin{align}
\psi(\tilde{x},\tilde{y},z) = \frac{1}{\big[
       z^2+\rho^2+a^2-2 a\rho
       \cos(\varphi +\Omega (t-t'))\big]^{1/2}
       -a\rho\Omega \sin(\varphi +\Omega(t-t'))}
\label{eq:LWSOLN}
\end{align}
as a concrete expression for \eqref{eq:MWresult}. With both 
Eqs.~\eqref{eq:seriesSOLN} and \eqref{eq:LWSOLN} at our 
disposal, we can evaluate the retarded solution 
to \eqref{eq:AppendixForcedHRWE} with enough accuracy 
(uniformly over the entire 2-center domain) to make 
the comparisons reported in Sec.~\ref{sec:numericaltests}.


\end{document}